\documentclass[10pt,journal,compsoc]{IEEEtran}
\IEEEoverridecommandlockouts
\usepackage{cite}
\usepackage[cmex10]{amsmath}
\usepackage{algorithm}
\usepackage{algorithmic}
\usepackage{array}
\usepackage{graphicx}
\usepackage{amssymb}
\usepackage{amsfonts}
\usepackage{mdwmath}
\usepackage{mdwtab}
\usepackage{eqparbox}
\usepackage{amsopn}
\usepackage{multirow}

\usepackage{subfig}
\usepackage{epstopdf}
\usepackage{extarrows}

\usepackage[dvips]{epsfig}
\usepackage{url}
\usepackage{xspace}
\usepackage{booktabs}
\graphicspath{{graphics/}}
\def\ie{\textit{i.e.}\xspace}
\def\etal{\textit{et al.}\xspace}
\def\etc{\textit{etc.}\xspace}

\def\eg{\textit{e.g.}\xspace}

\def\BuildSenSys{\textit{BuildSenSys}\xspace}
\def\BibTeX{{\rm B\kern-.05em{\sc i\kern-.025em b}\kern-.08em
    \kern-.1667em\lower.7ex\hbox{E}\kern-.125emX}}
\usepackage{color}

\begin{document}
\title{\BuildSenSys: Reusing Building Sensing Data for Traffic Prediction with Cross-domain Learning}

\author{Xiaochen~Fan, Chaocan~Xiang{*}, Chao~Chen, Panlong~Yang,~\IEEEmembership{Senior Member,~IEEE,}\\
Liangyi~Gong, Xudong~Song, Priyadarsi~Nanda and Xiangjian~He{*},~\IEEEmembership{Senior Member,~IEEE}
\IEEEcompsocitemizethanks{
\IEEEcompsocthanksitem Xiaochen Fan, Xudong Song, Priyadarsi Nanda and Xiangjian He are with the School of Electrical and Data Engineering, Faculty of Engineering and Information Technology, University of Technology Sydney, NSW 2007, Australia. E-mails: \{xiaochen.fan, xudong.song, priyadarsi.nanda, xiangjian.he\}@uts.edu.au.
\IEEEcompsocthanksitem Chaocan Xiang and Chao Chen are with the Key Laboratory of Dependable Service Computing in Cyber Physical Society (Chongqing University), Ministry of Education, China, and the College of Computer Science, Chongqing University, Chongqing 400044, China. E-mails: \{xiang.chaocan, ivanchao.chen\}@gmail.com.
\IEEEcompsocthanksitem Panlong Yang is with the School of Computer Science and Technology, University of Science and Technology of China, Hefei, Anhui 230026, China. E-mail: panlongyang@gmail.com.
\IEEEcompsocthanksitem Liangyi Gong is with the School of Software and BNRist, Tsinghua University, Beijing 100084, China. E-mail: gongliangyi@gmail.com.
\IEEEcompsocthanksitem Xiaochen Fan and Chaocan Xiang contributed equally to this work and share the first authorship. The corresponding authors are Chaocan Xiang and Xiangjian He.
}
}
\IEEEtitleabstractindextext{
\begin{abstract}
With the rapid development of smart cities,
smart buildings are generating a massive amount of building sensing data by the equipped sensors.
Indeed, building sensing data provides a promising way to enrich a series of data-demanding and cost-expensive urban mobile applications.
In this paper, as a preliminary exploration,
we study how to reuse building sensing data to predict traffic volume on nearby roads.
Compared with existing studies,
reusing building sensing data has considerable merits of cost-efficiency and high-reliability.
Nevertheless, it is non-trivial to achieve accurate prediction on such cross-domain data with two major challenges.
First, relationships between building sensing data and traffic data are not unknown as prior, and the spatio-temporal complexities impose more difficulties to uncover the underlying reasons behind the above relationships.
Second, it is even more daunting to accurately predict traffic volume with dynamic building-traffic correlations, which are cross-domain, non-linear, and time-varying.
To address the above challenges,
we design and implement \BuildSenSys, a first-of-its-kind system for nearby traffic volume prediction by reusing building sensing data.
Our work consists of two parts, \ie, \emph{Correlation Analysis} and \emph{Cross-domain Learning}.
First, we conduct a comprehensive building-traffic analysis based on multi-source datasets, disclosing \emph{how} and \emph{why} building sensing data is correlated with nearby traffic volume.
Second, we propose a novel recurrent neural network for traffic volume prediction based on cross-domain learning with two attention mechanisms.
Specifically, a cross-domain attention mechanism captures the building-traffic correlations and adaptively extracts the most relevant building sensing data at each predicting step.
Then, a temporal attention mechanism is employed to model the temporal dependencies of data across historical time intervals.
The extensive experimental studies demonstrate that \BuildSenSys outperforms all baseline methods
with up to 65.3\% accuracy improvement (\eg, 2.2\% MAPE) in predicting nearby traffic volume.
We believe that this work can open a new gate of reusing building sensing data for urban traffic sensing,
thus establishing connections between smart buildings and intelligent transportation.
\end{abstract}

}

\maketitle

\section{Introduction}\label{introduction}
\begin{figure}[t]
	\centering
	\includegraphics[width=0.45\textwidth]{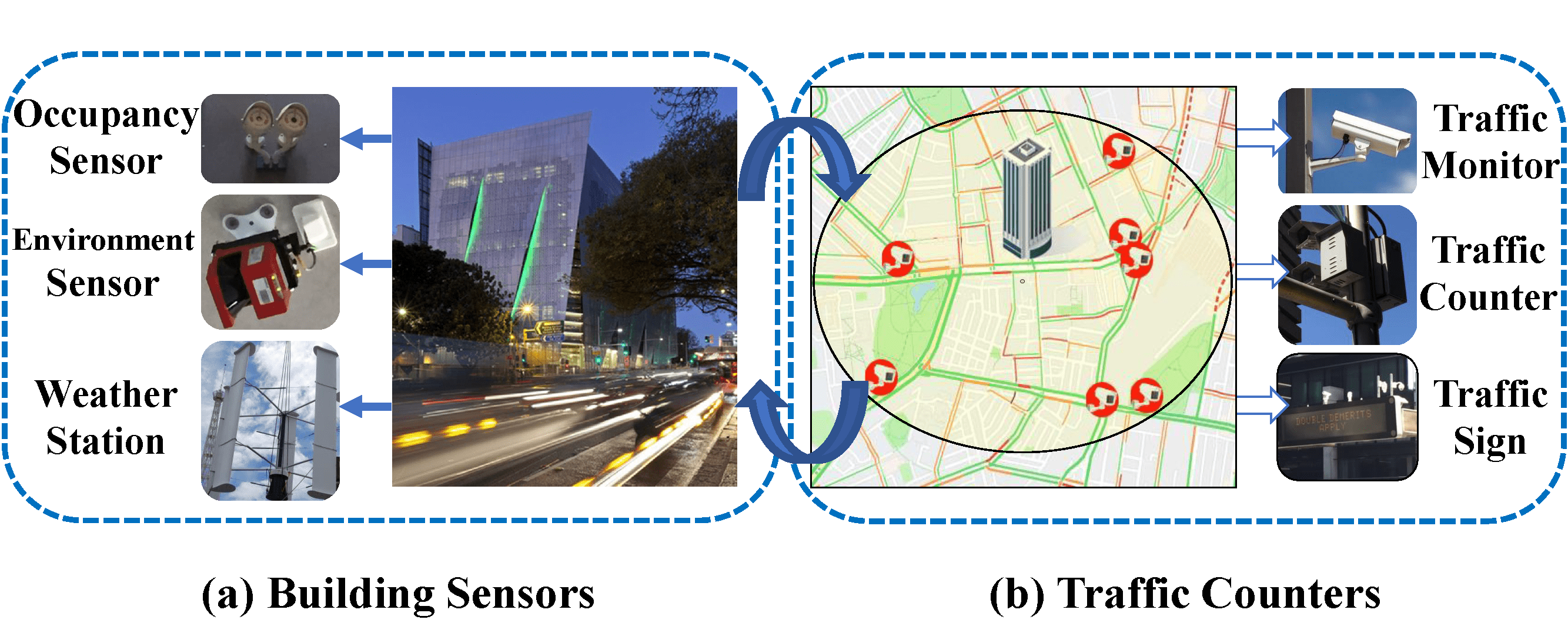}
	\caption{Illustrations for reusing building sensing data to predict nearby traffic volume with cross-domain learning.}
	\label{introexample}
\end{figure}

\IEEEPARstart{S}{mart} buildings equipped with an increasing number of IoT sensors are rising rapidly,
thus producing large amount of building sensing data (also called \emph{building data}\footnote{In the remaining of this paper, we will use the terms ¡®building sensing data¡¯ and ¡®building data¡¯ interchangeably unless otherwise stated.}
~\cite{plageras2018efficient,bates2017beyond,nesa2017iot}).
According to Statista's reports in~\cite{SmartBuildingsData2019},
the explosive volume of sensing data collected by global smart buildings was nearly 7.8 ZB~(about $7.8\times 2^{40}$G) in 2015, and it is expected to be growing up to nearly 37.2 ZB by 2020.
For instance, as shown in Fig.~\ref{introexample}(a),
a CBD building in Sydney is installed with more than 2,000 sensors, generating over 100 million sensor readings in
monitoring the status of the building,
including building occupancy, indoor/outdoor environment, \etc

Reusing existing building sensing data is with great significance for ubiquitous sensing of smart cities, as building data is cost-efficiency and sustainable~\cite{Snoonian2003IEEE}.
Assuming building sensing data is available for the long-term,
it has recently enabled a series of new applications in urban sensing,
such as developing urban 3D mobility models~\cite{zheng2018buildings}
and identifying abnormal appliances~\cite{rashid2018, zheng2016urban}.
In this paper, as a preliminary exploration (illustrated in Fig.~\ref{introexample}),
we are motivated to reuse building sensing data for nearby traffic sensing for the following reasons. Intuitively, most urban buildings are connected by roads,
while residents move among different buildings, mainly via commuting on these roads~\cite{jo2018image}.
According to the report in~\cite{klepeis2001national},
the American citizens averagely spend over 93\% of their daily time in enclosed buildings and vehicles (accounting for 87\% and 6\%, respectively).
Thus, there could exist some underlying relationships between sensing data of buildings and traffic data of buildings' surrounding roads,
named as \emph{building-traffic correlations}~\cite{zheng2016urban}.
Such relationships reveal a promising direction of reusing building sensing data,
\ie, for predicting traffic of nearby roads.
As a proof-of-concept, this work mainly focuses on predicting \emph{traffic volume, which is defined as the number of vehicles traversing on a road segment per hour}~\cite{zhan2017citywide}.
Accurate predictions on traffic volume are fundamentally crucial for Intelligent Transportation Systems~(ITSs), such as traffic light control, road navigation, and estimation of vehicle emission~\cite{Deng2016KDD,zheng2016urban,nagy2018survey}.

Compared with the existing methods~\cite{tang2015hybrid,ide2016city,zhan2017citywide,lin2018road,wan2016mobile,qin2018eximius} of predicting nearby traffic volume, reusing building data has considerable merits of both \emph{low cost} and \emph{high reliability},
which are crucial to traffic sensing~\cite{nagy2018survey}.
Most of the conventional traffic prediction methods heavily rely on the fixed road-based traffic sensing systems, such as loop detectors~\cite{tang2015hybrid} and traffic surveillance cameras~\cite{ide2016city}.
Although these methods can provide accurate traffic volume information,
they incur extremely high costs on installation and maintenance,
impeding their extensions to large-scale cities~\cite{zheng2016urban}.
On the contrary,
re-using off-the-shelf building sensing data will significantly reduce the cost,
as with no extra deployment and maintenance~\cite{zheng2016urban,zhan2017citywide}.
In terms of opportunistic sensing data of floating vehicles,
GPS trajectories~\cite{zhan2017citywide} and cellular records of passengers~\cite{qin2018eximius} are recently exploited to infer traffic volume
to address the high-cost issue.
However, opportunistic sensing suffers from the uncontrollable property of users and data deficiency of particular roads, resulting in unreliable performance in traffic volume prediction~\cite{qin2018eximius}.
Buildings and their surrounding roads,
as two basic constructions in urban infrastructure,
are both stationary with long-term spatial relations.
Therefore, in comparison to the opportunistic sensing~\cite{zhan2017citywide},
buildings can generate more reliable and sustainable sensing data,
providing opportunities for accurately predicting nearby traffic volume for long term~\cite{balaji2016brick,wan2016mobile}.

Nevertheless, it is necessary to formally address two principal challenges in achieving reusing building sensing data for predicting nearby traffic volume.



\begin{itemize}
\item \textbf{Challenge 1}: \emph{Investigating unknown building-traffic relationships}. While a few works~\cite{zheng2016urban,zheng2018buildings} have shown certain evidence of building-traffic correlations, they failed to investigate what exactly these relationships are.
    To precisely reveal building-traffic correlations is difficult with great complexity.
    Furthermore, it is extremely challenging to uncover the underlying reasons behind these correlations.
\item \textbf{Challenge 2}: \emph{Accurate prediction with cross-domain building-traffic correlations}. The building-traffic relationships are cross-domain and non-linear, which further renders accurate prediction of traffic volume as non-trivial. Even more challenging, these correlations would vary dynamically along with time, imposing more difficulties with accurate traffic prediction~\cite{pan2019urban}.
\end{itemize}

To address these two challenges, we design and implement \textbf{\BuildSenSys},
a first-of-its-kind \textbf{\emph{Build}}ing \textbf{\emph{Sen}}sing data-based \textbf{\emph{Sys}}tem for nearby traffic volume prediction.
First, for Challenge 1, we conduct extensive experiments based on multi-source real-world datasets (in Section~\ref{observation}).
We delve into the relationships between building data and nearby traffic volume, and the experimental results indicate that building-traffic correlations are
\emph{non-linear}, \emph{time-varying}, and \emph{cross-domain}.
Second, for Challenge 2, we propose a novel recurrent neural network based on cross-domain learning with two attention mechanisms for traffic volume prediction (in Section~\ref{architecture}).
Specifically, a cross-domain attention mechanism captures the building-traffic correlations and adaptively extracts the most relevant building data at each time interval.
Then, a temporal attention mechanism is employed to model the temporal dependencies across historical time intervals.
Finally, we implement a prototype system of \BuildSenSys
and conduct comprehensive evaluations with one-year real-world datasets.

In summary, we make three key contributions in this paper as follows:

\begin{itemize}
\item To the best of our knowledge, we are the first to conduct comprehensive building-traffic analysis with multi-source real-world datasets. By applying multi-source cross-verification, this work not only discloses \emph{how} but also sheds light on \emph{why} the building data is correlated with nearby traffic volume.
    Essentially, the changes in building occupancy induced by the commuting activities of occupants are highly related to the dynamics of traffic volume on nearby roads. In addition, the higher probability building occupants pass through a road, the stronger building-traffic correlation there exists.

\item We propose a novel recurrent neural network for traffic volume prediction based on cross-domain learning. It leverages a cross-domain attention-based encoder and a temporal attention-based decoder to extract the \emph{non-linear}, \emph{time-varying}, \emph{cross-domain} building-traffic correlations accurately and further achieves accurate traffic volume prediction.

\item The extensive experimental studies demonstrate that \BuildSenSys outperforms all baseline methods with up to 65.3\% accuracy improvement (\eg, 2.2\% mean absolute percentage error) in predicting nearby traffic volume. We believe this work can open a new gate of reusing building sensing data for traffic sensing, and further establish connections between smart buildings and intelligent transportation.
\end{itemize}

The rest of this paper is organized as follows.
Section~\ref{related_work} reviews related literature and Section~\ref{framework} introduces an overview of the \BuildSenSys system.
Then, Section~\ref{observation} presents the correlation analysis of building data and nearby traffic data.
Section~\ref{architecture} formulates the prediction problem and presents
the cross-domain learning-based recurrent neural network for traffic prediction.
Section~\ref{simulation} evaluates \BuildSenSys through extensive experimental studies with real-world datasets.
Finally, Section~\ref{discussion} discusses some critical issues of reusing building data,
and Section~\ref{conclusion} concludes the paper.

\section{Related Work}\label{related_work}
{\textbf{Reusing building data}}:
Building sensing data is originally dedicated to management and control purposes~\cite{mohanty2016everything};
therefore, reusing building sensing data for traffic prediction will not cost any extra cost.
Beyond building management, reusing building sensing data has attracted considerable research interests from smart city applications, including urban transportation~\cite{zheng2016urban}, crowd flow patterns~\cite{zheng2018buildings}, and data integration~\cite{minoli2017iot}.
According to~\cite{liu2016mining}, a single set of traffic monitoring systems with camera detectors could easily cost \$2500 USD, and over 100 million dollars of such devices can only cover a quarter of roads in a typical metropolitan city~\cite{zheng2016urban}.
In comparison with conventional methods~\cite{ide2016city,zhan2017citywide,lin2018road,wan2016mobile}
that fully rely on sensing data from traffic monitoring systems,
it is considerably low-cost while highly reliable to make a second use of building sensing data to predict nearby traffic.
For example, Zheng~\etal\cite{zheng2018buildings} studied the impacts of buildings on human movements and further developed a new urban mobility model for urban planning.
Hu~\etal\cite{hu2018stube+} proposed a communication sharing architecture for smart buildings to organize in-building IoT devices with heterogeneous data communication.
Moreover, Zheng~\etal\cite{zheng2016urban} proposed to use indoor ${\rm{C}}{{\rm{O}}_2}$ data to estimate building occupancy and further developed an occupancy-traffic model for traffic speed prediction.
Inspired by the above existing works,
it is of great significance to unify traffic monitoring systems with external sensing infrastructures to enhance traffic prediction accuracy and reduce marginal cost.
In this study, we propose innovative reuse of multi-dimension building sensing data for traffic prediction.
As a preliminary exploration, we extensively investigate the building-traffic correlations and apply cross-domain learning to achieve accurate traffic prediction.

\textbf{Traffic volume prediction}:
Most existing works of predicting traffic volume use data from fixed road-based traffic sensors, such as loop detectors, microwave radars, and video cameras~\cite{nagy2018survey}.
The main advantage of road-based sensors is that they can provide reliable data by capturing all vehicles passing by the corresponding roads~\cite{nellore2016survey}.
Recently, opportunistic sensing and crowdsensing techniques have been utilized to collect GPS data~\cite{zhan2017citywide,kong2018lotad,liu2018think}, and cellular record data~\cite{cui2017mining,janecek2015cellular,costa2017towards} from floating vehicles and mobile passengers.
These trajectories provide detailed mobility traces for
network-wide traffic sensing and prediction.
Many existing works have integrated both traffic sensor data and opportunistic sensing data for traffic prediction.
For example, Meng~\etal\cite{meng2017city} further combined loop detector data and taxi trajectories with a spatio-temporal semi-supervised learning model to infer traffic volume.
However, with the inherent biases and random data deficiency,
most trajectory data cannot cover the entire traffic dynamics.
For instance, GPS data of a 6,000-taxi network can only cover 28\% of the overall road segments in a large city with distinctive operating time~\cite{qin2018eximius}.
With such data sparsity, trajectory sensing data cannot guarantee the sustainability and reliability in traffic prediction,
especially on explicitly targeted road segments.
In this work,
the source data for traffic prediction is building sensing data,
which is quite different from most existing works.
Both the buildings and their surrounding roads are stationary,
and inherently they have permanent spatial relations.
Most importantly, as building-traffic correlations are sustainable,
reusing building data is cost-efficiency and highly reliable for long-term traffic prediction.

\textbf{Traffic prediction models}:
Conventional short-term traffic prediction approaches mainly apply parameter-based prediction models,
for example, Autoregressive Integrated Moving Average (ARIMA)~\cite{sarker2018morp},
Vector Autoregression (VAR)~\cite{pavlyuk2017short} and Locally Weighted Linear Regression (LWR)~\cite{zheng2016urban}.
With the rising of deep learning,
deep neural networks~\cite{lv2015traffic}have been adopted for traffic prediction as they can capture complex temporal-spatial dependencies through feature learning~\cite{liu2018urban}.
Moreover, Recurrent Neural Networks (RNN) have also been adopted by~\cite{xu2018real} to perform sequence learning on historical traffic data.
However, RNNs are not capable of preserving long-term dependencies on historical traffic data,
as their performance would deteriorate with longer input.
As a result, Long Short-Term Memory (LSTM) networks are further adopted by many research studies~\cite{zhao2017lstm,tian2018lstm} to perform long-term prediction tasks.
Inspired by the human's ability to capture a focus in certain visions,
attention mechanisms have been integrated into the neural networks for sequence-to-sequence learning~\cite{vaswani2017attention}.
For instance, \cite{yao2019revisiting} proposed a spatial-temporal dynamic network with a periodically shifted attention mechanism
to capture periodic temporal similarity in traffic predictions.
In this paper, we devise a cross-domain learning-based recurrent neural network with
a cross-domain attention mechanism and a temporal attention mechanism.
Our model can effectively extract cross-domain, non-linear, and time-varying building-traffic correlations for accurate traffic prediction.

\begin{figure}[t]
	\centering
	\includegraphics[width=0.495\textwidth]{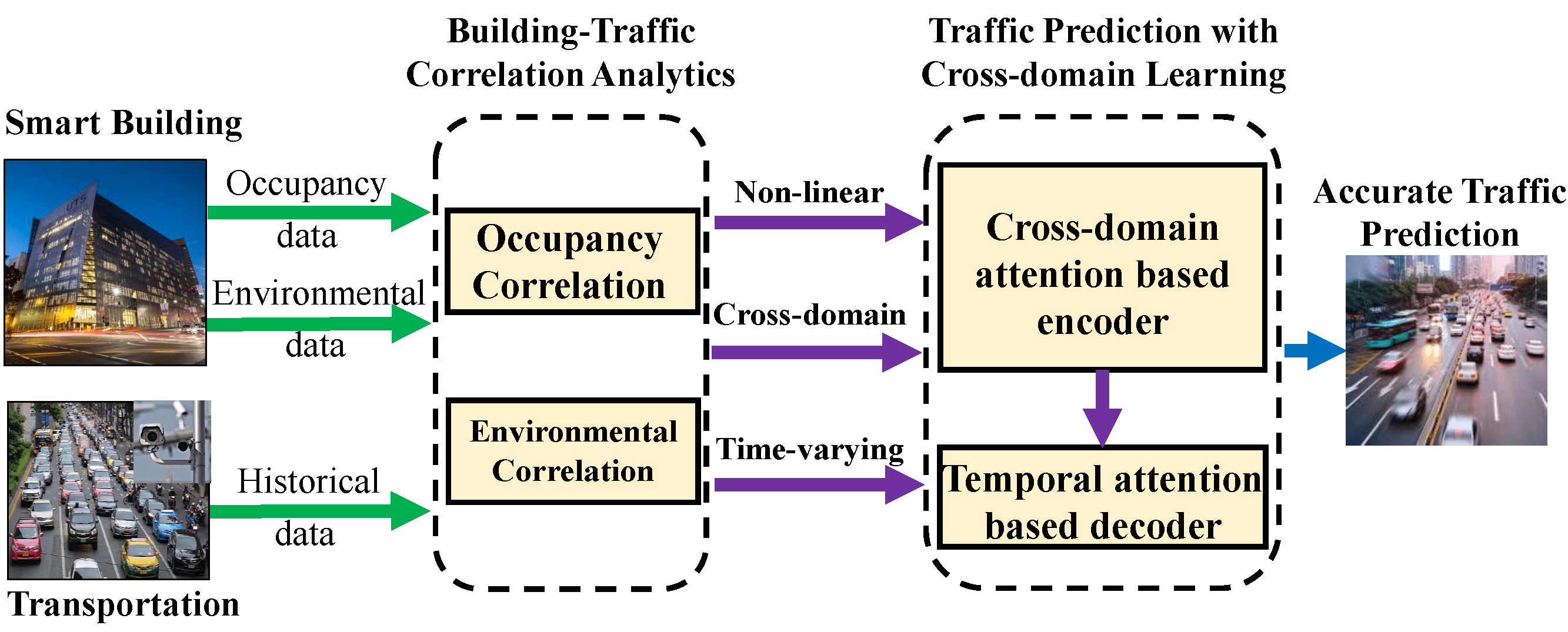}
	\caption{An overview of the BuildSenSys system for reusing building data for nearby traffic prediction.}
	\label{system_architecture}
\end{figure}

\begin{figure*}[t]
	\centering
	\subfloat [Locations]
	{
		\label{traffic_location}
		\begin{minipage}[t]{0.15\linewidth}
			\centering
			\includegraphics[width=0.99\textwidth]{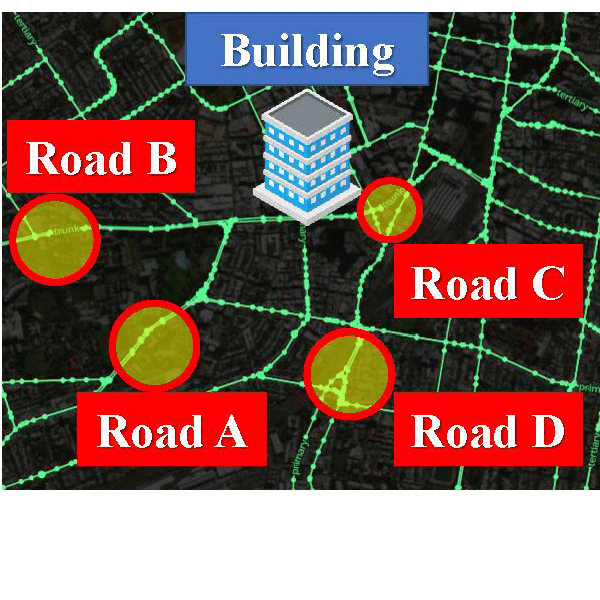}
			\par\vspace{0pt}
		\end{minipage}
	}
	\subfloat [Road A]
	{
		\label{comparison_a}
		\begin{minipage}[t]{0.21\linewidth}
			\centering
			\includegraphics[width=0.99\textwidth]{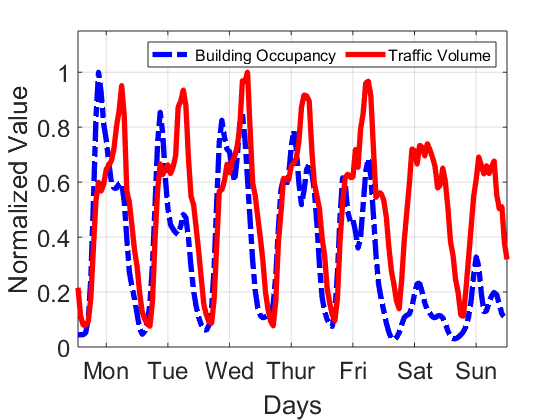}
			\par\vspace{0pt}
		\end{minipage}
	}
    \subfloat [Road B]
	{
		\label{comparison_b}
		\begin{minipage}[t]{0.21\linewidth}
			\centering
			\includegraphics[width=0.99\textwidth]{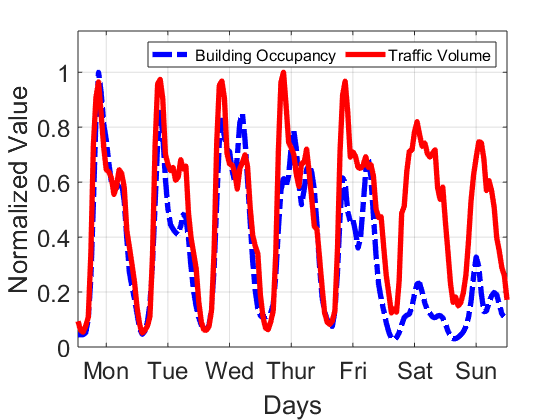}
			\par\vspace{0pt}
		\end{minipage}
	}
    \subfloat [Road C]
	{
		\label{comparison_c}
		\begin{minipage}[t]{0.21\linewidth}
			\centering
			\includegraphics[width=0.99\textwidth]{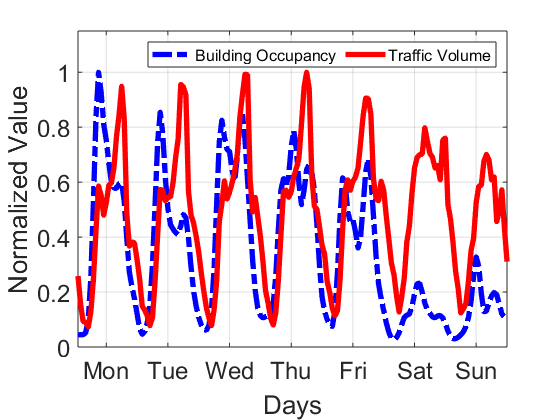}
			\par\vspace{0pt}
		\end{minipage}
	}
    \subfloat [Road D]
	{
		\label{comparison_d}
		\begin{minipage}[t]{0.21\linewidth}
			\centering
			\includegraphics[width=0.99\textwidth]{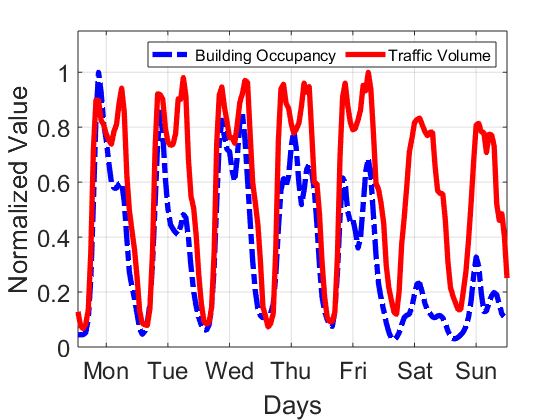}
			\par\vspace{0pt}
		\end{minipage}
	}
	\caption{Comparisons of normalized building occupancy and normalized traffic volume on different roads.}
	\label{traffic_map}
\end{figure*}

\section{System Overview}\label{framework}
In this section, we briefly present the system overview of reusing building data for traffic volume prediction.
As shown in Fig.~\ref{system_architecture},
the data sources of \BuildSenSys include a smart building (generating occupancy data and environmental data) and nearby traffic monitors (providing traffic volume data).
More importantly, \BuildSenSys consists of two main components as follows.

1) \textbf{Building-traffic correlation analysis with multi-source real-world datasets}~(in Section~\ref{observation}).
Extensive experimental analysis is conducted based on real-world smart building data, traffic data of fixed-road sensors, and Google traffic data.
We investigate not only whether building data is correlated
with the traffic data but also what the underlying reasons are.
The results show that the answer is positive, and there are two types of cross-domain correlations, including
i)~\textbf{occupancy correlation} (in Section~\ref{SpaAnalysis}) between the building occupancy and traffic data,
and ii)~\textbf{environmental correlation} (in Section~\ref{EnvAnalysis}) between the building environmental data and traffic data.
Besides, the building-traffic correlations are dynamic and time-varying, generating the temporal correlations along with the time.

2) \textbf{Cross-Domain Learning for Accurate Traffic Prediction}~(in Section~\ref{architecture}).
In correspondence to cross-domain correlations and temporal correlations,
we propose cross-domain learning to predict traffic volume accurately.
Specifically, we propose a cross-domain attention mechanism for the encoder to learn features from non-linear,
cross-domain correlations between building data and traffic data (in Section~\ref{Encoder-method}).
By correlating these features with traffic volume data,
the most relevant cross-domain correlations are extracted through the training process of the \BuildSenSys model.
Furthermore, we present a temporal attention mechanism for the decoder,
which aims to learn the temporal dependencies for the predicted traffic data (in Section~\ref{Decoder-method}). After that, the output context factors are leveraged to predict traffic volume.

\section{Building-traffic Correlation Analysis with Multi-source Datasets}\label{observation}
In this section, we conduct comprehensive experiments to explore spatial and temporal building-traffic correlations with multi-source real-world datasets as follows. To begin with, we briefly introduce the basics of building sensing data and traffic volume data as follows.

First, we collect building sensing data from the Faculty of Engineering and Information Technology's Building (FEIT Building) at the University of Technology Sydney,
New South Wales, Australia.
The FEIT building is a 16-level campus building with a total usable floor area of 23,500 ${m^2}$
and it has the capacity to accommodate a maximum population of over 5000.
As illustrated in Fig.~\ref{introexample}(a),
the FEIT building is described as a `living laboratory,'
with around 2,500 internal environment sensors installed across all floors and public spaces.
These sensors constantly monitor the internal and external environments of the building,
including indoor environment (by environmental sensors), outdoor environment (by a roof-top weather station), and building occupancy (by smart cameras).
We access all building sensing data through an online database server via MySQL workbench,
which contains 33 types of building sensing data with a total volume of over 10 GB~\cite{EIF_Wiki}.
In this paper, we have leveraged 10 most relevant building sensing data from FEIT building to achieve cross-domain traffic sensing and prediction.

Second, as shown in Fig.~\ref{introexample}(b), the traffic data are the traffic volumes collected by permanently deployed traffic counters on several road segments
that are proximate to the FEIT building.
Each traffic counter is integrated to a traffic monitoring system for calculating hourly traffic volume data.
We access historical traffic volume data from the official website of the Department of Roads and Maritime Services, New South Wales State~\cite{nsw2018traffic}.
In the following, we present a detailed cross-domain correlation analysis with traffic volume data and different building sensing data.

\subsection{Correlation analysis with building occupancy data}\label{SpaAnalysis}
In this section, we investigate cross-domain correlations between building occupancy data and traffic volume.
The building occupancy data is collected by a number of cameras
installed at building entrances, stairways, and walkways to monitor the people's movement inside the building.
Based on the sensing data of these cameras,
we use the PLCount algorithm~\cite{sangoboye2016plcount} to compute the total building occupancy accurately.
As the calculation of building occupancy is not our focus, we omit its details that can be referred from~\cite{sangoboye2016plcount}.


\begin{figure*}[t]
	\centering
	\subfloat [Different roads~(Cosine)]
	{
		\label{Correlation:subfig_a}
		\begin{minipage}[t]{0.24\linewidth}
			\centering
			\includegraphics[width=0.99\textwidth]{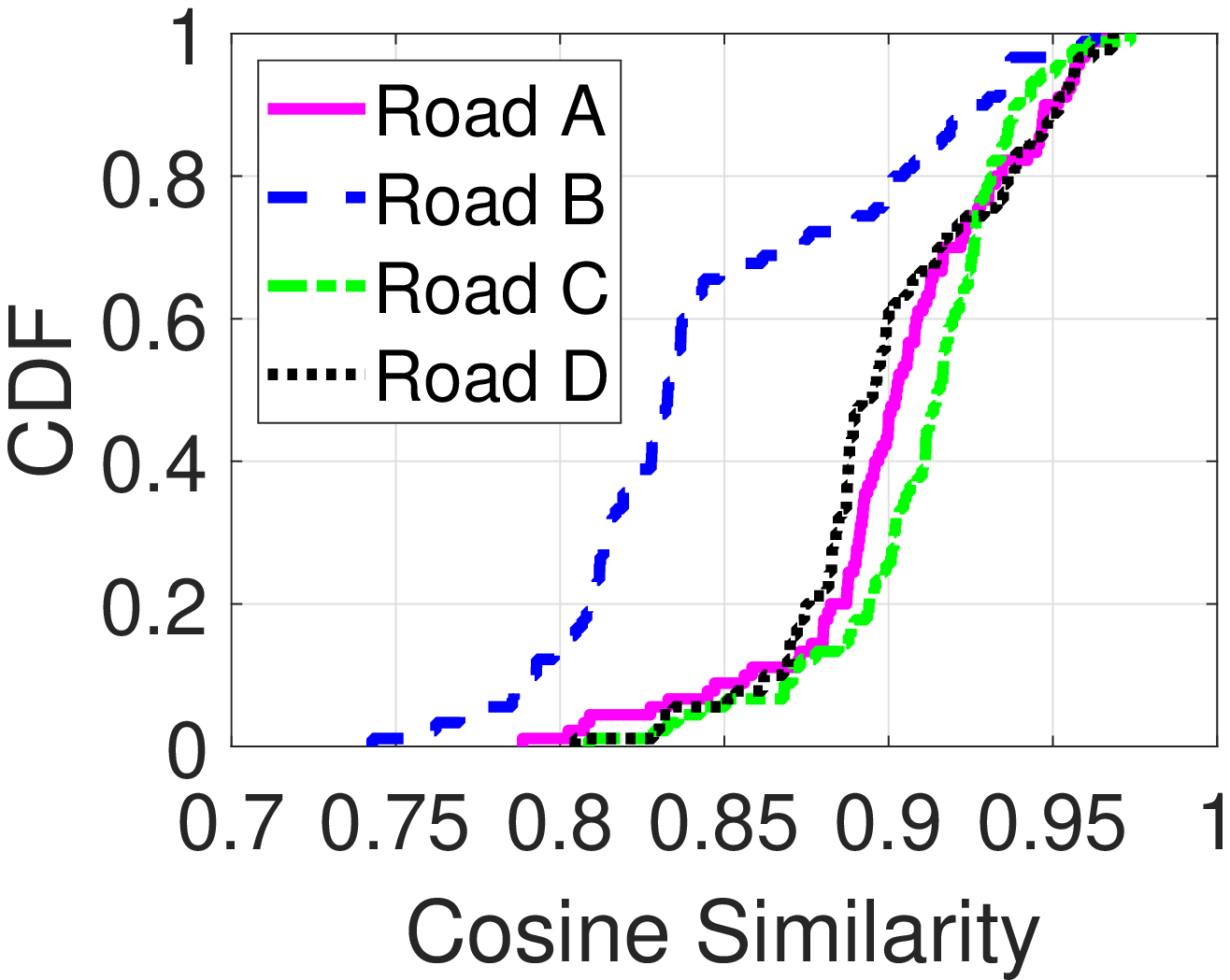}
			\par\vspace{0pt}
		\end{minipage}
	}
	\subfloat [Different roads~(Pearson)]
	{
		\label{Correlation:subfig_b}
		\begin{minipage}[t]{0.24\linewidth}
			\centering
			\includegraphics[width=0.99\textwidth]{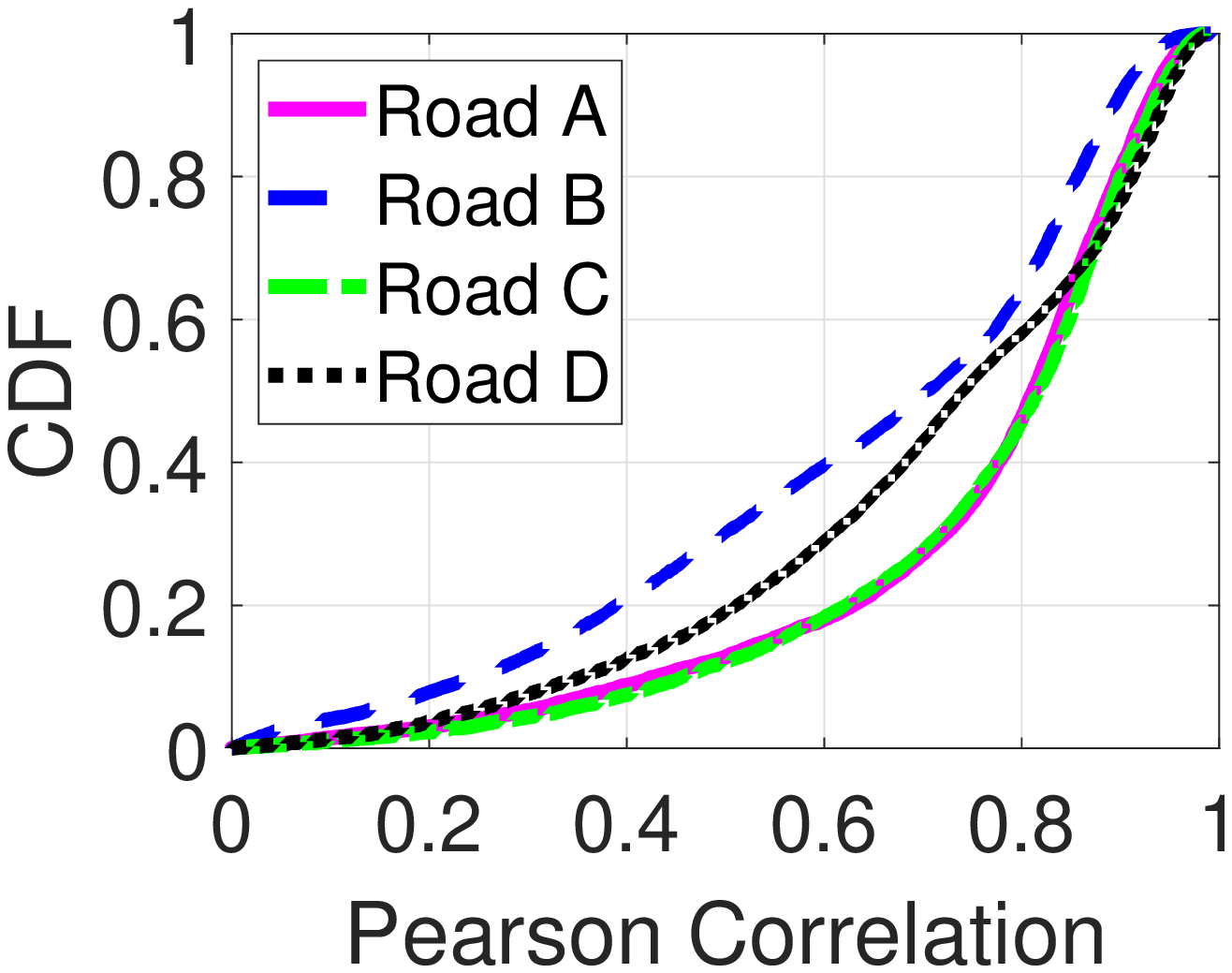}
			\par\vspace{0pt}
		\end{minipage}
	}
	\subfloat [Weekday vs. Weekend]
	{
		\label{Correlation:subfig_c}
		\begin{minipage}[t]{0.24\linewidth}
			\centering
			\includegraphics[width=0.99\textwidth]{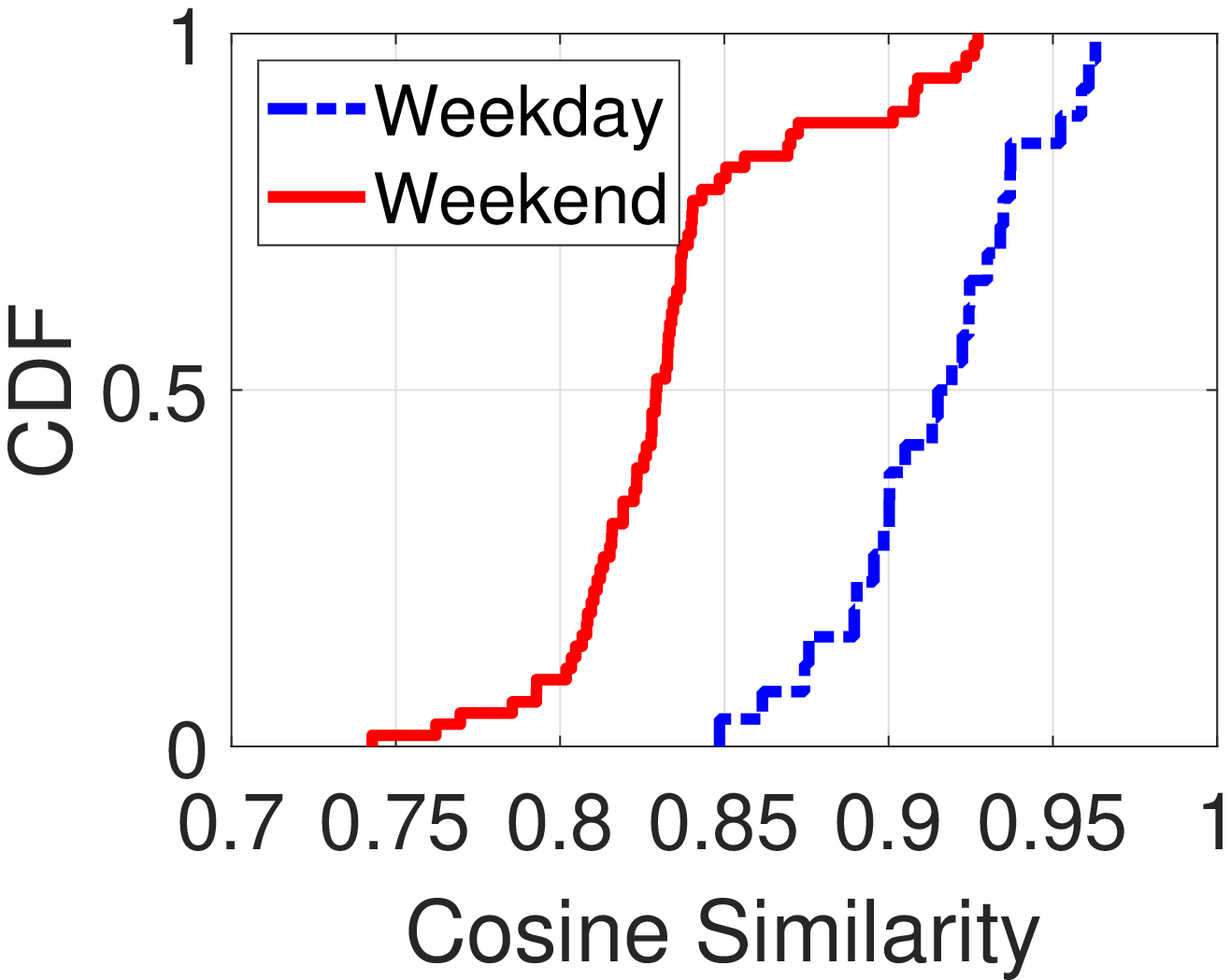}
			\par\vspace{0pt}
		\end{minipage}
	}
	\subfloat [Normal hour vs. Rush hour]
	{
		\label{Correlation:subfig_d}
		\begin{minipage}[t]{0.24\linewidth}
			\centering
			\includegraphics[width=0.99\textwidth]{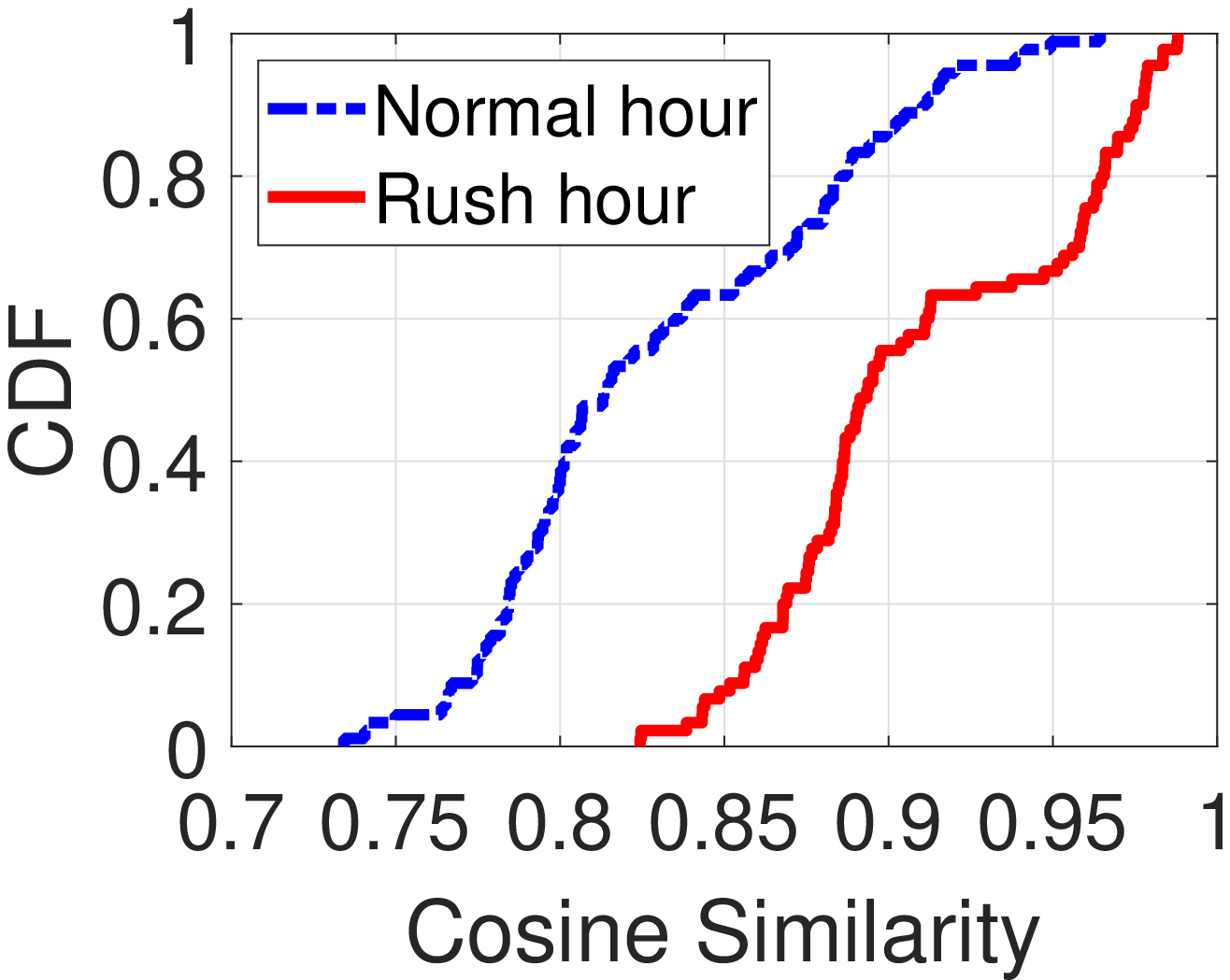}
			\par\vspace{0pt}
		\end{minipage}
	}
	\caption{Quantification of correlations between building occupancy data and traffic volume data with two metrics.}
	\label{Fig_CorrelationWeekdayVSWeekend} 
\end{figure*}

\subsubsection{Correlation quantification with metrics}
\label{subsec:Correlationquantify}
With the preprocessed building occupancy,
we closely study the correlations between building occupancy and nearby traffic volume.
First, we compare building occupancy with the traffic volume of nearby roads in one week.
Fig.~\ref{traffic_map} shows the comparison of normalized building occupancy (dashed blue line) and traffic volume (solid red line) of roads A-D.
Note that the above four roads are different types of roads,
where road A is a minor highway, road B is a major highway, road C is a main street,
and road D is a primary street.
We observe that the building occupancy and traffic volume exhibit similar hourly patterns on weekdays and different patterns on weekends, respectively.
This is because the high volume of building occupancy is driven by the working days, and there are just a few overtime workers inside the building on weekends.
In particular, the building occupancy closely follows the rise of traffic volume in both morning rush hours and evening rush hours.

However, the comparison between normalized building occupancy and traffic volume only shows the general dynamics and correlations.
To further quantify the correlations between building occupancy and traffic volume,
we employ two correlation metrics, \ie, Cosine Similarity and Pearson Correlation Coefficient,
to evaluate the above correlations.
The \emph{Cosine Similarity} quantifies the similarity between two non-zero vectors on an inner product space.
We use ${s_i}$ to represent the cosine similarity between building occupancy and traffic volume on the day $i$.
It is computed as ${s_i} = \cos ({{\bf{b}}_i},{{\bf{t}}_i})$,
where ${{\bf{b}}_{\bf{i}}} = [{b_{i,1}},{b_{i,2}},...,{b_{i,24}}]$ and ${{\bf{t}}_{\bf{i}}} = [{t_{i,1}},{t_{i,2}},...,{t_{i,24}}]$ denote the vectors of building occupancy and traffic volume over 24 hours on the day $i$, respectively.
In this way, the overall cosine similarity between building occupancy and traffic volume for $n$ days is denoted by
${\bf{s}} = [{s_1},...{s_i},...{s_n}]$.
Moreover, the \emph{Pearson Correlation} is computed as
$p = \sum\limits_{j = 1}^k {({b_{i,j}} - {{\bar b}_i})({t_{i,j}} - {{\bar t}_i})} /\sqrt {\sum\limits_{j = 1}^k {{{({b_{i,j}} - {{\bar b}_i})}^2}\sum\limits_{j = 1}^k {{{({t_{i,j}} - {{\bar t}_i})}^2}} } }$,
where $k$ denotes the time interval for each Pearson Coefficient,
${\bar b_i}$ and ${\bar t_i}$ are the average building occupancy and traffic volume at the day $i$, respectively.
Note that the absolute value of Pearson Correlation indicates the strength of the correlation,
ranging from 0 (weak correlation) to 1 (strong correlation).




We quantify correlations between building occupancy and traffic volume on four different road segments, using both Cosine Similarity and Pearson Correlation.
As illustrated in Fig.~\ref{Correlation:subfig_a}, the cosine similarities on all four roads are greater than 0.8, showing that building occupancy is strongly correlated with traffic volume. Meanwhile, Fig.~\ref{Correlation:subfig_b} shows that over 70\% of traffic volume data has strong and positive ($\ge$0.5) Pearson Correlation with building occupancy data.
In addition, we evaluate the cosine similarity of building occupancy and traffic volume by dividing the data into different groups, \ie, weekdays/weekends and rush hours/normal hours, respectively.
From Fig.~\ref{Correlation:subfig_c} and Fig.~\ref{Correlation:subfig_d}, it is clear that building occupancy and traffic volume have higher correlations on the weekdays ($\ge$0.85) and rush hours ($\ge$0.82) than the weekends ($\ge$0.75) and normal hours ($\ge$0.75), respectively.

\begin{figure}[t]
	\centering
	\begin{minipage}[t]{0.9\linewidth}
	\centering
	\includegraphics[width=0.99\textwidth]{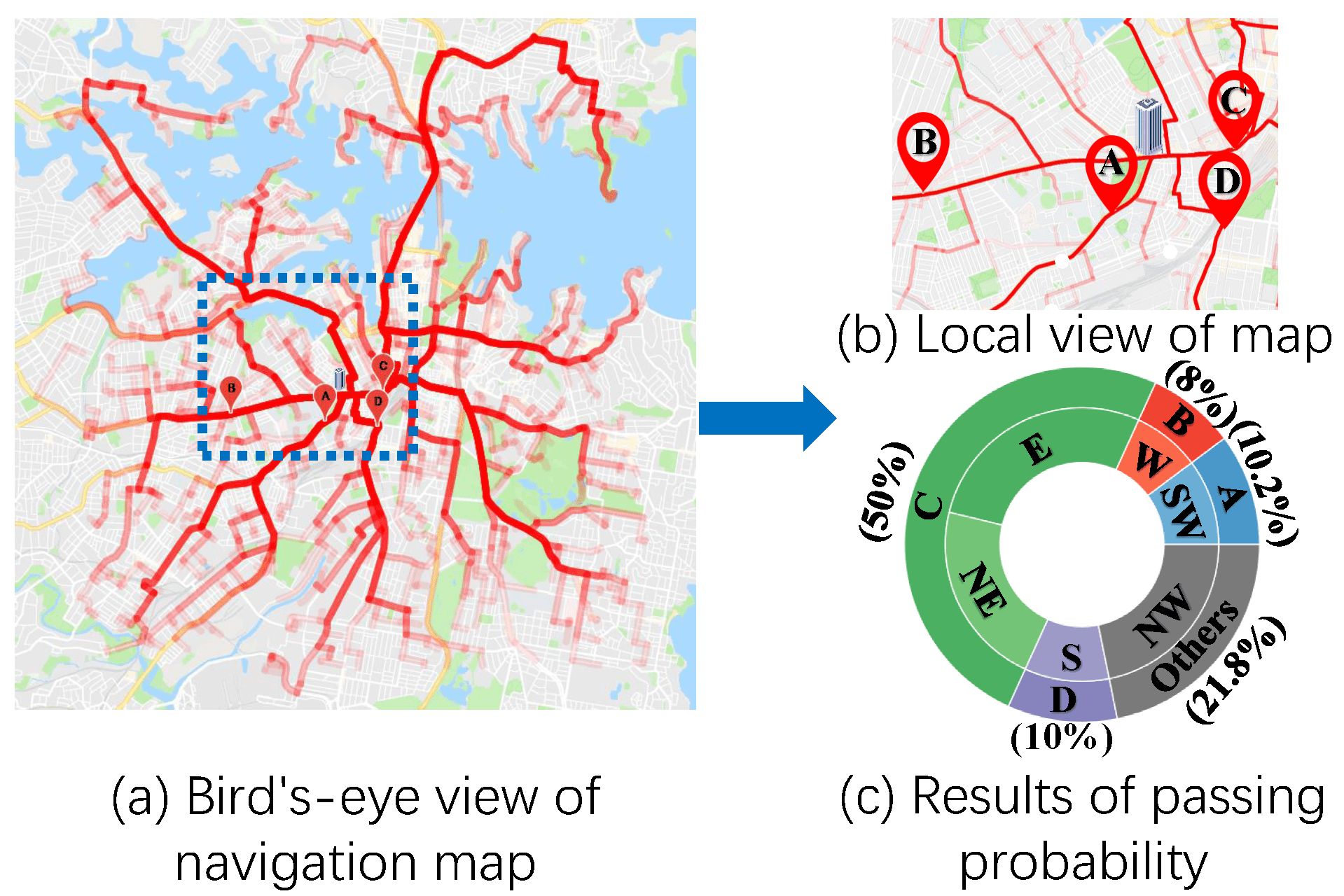}
	\par\vspace{0pt}
	\end{minipage}
	\caption{Cross-verification for building-traffic correlations via Google Maps Navigation.}
	\label{Fig_GoogleApi} 
\end{figure}

\subsubsection{Correlation verification with Google Maps}
The results in Section~\ref{subsec:Correlationquantify} indicate that the building occupancy data is highly correlated with the traffic volume on nearby road segments.
Through an in-depth analysis, the main reason is that
most of the building occupants entering or leaving a building would pass the surrounding roads via transportation.
Intuitively, the building-traffic correlations are dependent on the probability of building occupants passing by each road segment.
That is, the building-traffic correlation would be stronger when a road is with higher passing probability by the building's occupants.

To verify this assumption,
we further investigate the relationship between building occupancy and traffic volume by conducting cross-verification experiments with Google Maps.
First, as shown in Fig.~\ref{Fig_GoogleApi}(a), we set this building's location as the starting points or the endpoints, while randomly selecting 500 locations within 10 km distance of this building as the end/starting points.
Each randomly selected point together with the building's location are packaged as a navigation request,
which is sent to Google Maps for navigation.
Then, the best-fit road geometries are returned with a series of GPS locations.
We visualize 500 navigation results in Fig.~\ref{Fig_GoogleApi},
where the color of each route indicates its passing probability and a route with deeper color has a higher passing probability.

As illustrated in Fig.~\ref{Fig_GoogleApi}(a), some long-distance routes would pass through the main roads, for example, major highways and primary streets.
Moreover, Fig.~\ref{Fig_GoogleApi}(b) shows a zoom-in view of roads A-D. Specifically, the passing probabilities of roads A-D are shown in Fig.~\ref{Fig_GoogleApi}(c), where the outer circle represents the passing probability of four road segments, and the inner circle represents the main directions of navigation routes.
It can be observed that the passing probabilities of roads A, B, C, D are 10.2\%, 8\%, 50\% and 10\%, respectively, while the others only have about 20\% passing ratio.
By using the cross-verification method, we compare these results based on Google Maps with building-traffic correlations in Section~\ref{subsec:Correlationquantify}.
As indicated in Table~\ref{ComparisonResults}, the results from multi-source datasets are consistent with each other.
That is, a road segment with higher passing probability and shorter distance to the building would have stronger building-traffic correlations with this building.
As a result, the above results of cross-verification further verify the reason behind building-traffic correlation as follows. \emph{The building occupants entering or leaving a building would pass the surrounding roads via transportation. Hence, the change of building occupancy data (induced by the commuting activity of the building's occupants) is highly related to the traffic volume of nearby road segments.}

\begin{table}[t]
\begin{tabular}{l|l|l|l|l}
\hline
Road Name & Passing Ratio & Cosine      & Pearson  & Distance                 \\ \hline\hline
Road C    & 50\%          & 1st         & 1st      & 0.3 km                   \\ \hline
Road A    & 10.2\%        & 2nd         & 2nd      & 1.6 km                   \\ \hline
Road D    & 10\%          & 3rd         & 3rd      & 1.9 km                   \\ \hline
Road B    & 8\%           & 4th         & 4th      & 3.0 km                   \\ \hline

\end{tabular}
\caption{Cross-verification: comparison between roads A, B, C and D in navigation passing probability, Cosine similarity, Pearson correlation and distance to the building.}
\label{ComparisonResults}
\end{table}

\subsection{Correlation analysis with environmental data}\label{EnvAnalysis}

\begin{figure}[t]
\centering
\subfloat [Outdoor AQI]
	{
		\label{aqi_traffic}
		\begin{minipage}[t]{0.48\linewidth}
			\centering
			\includegraphics[width=0.99\textwidth]{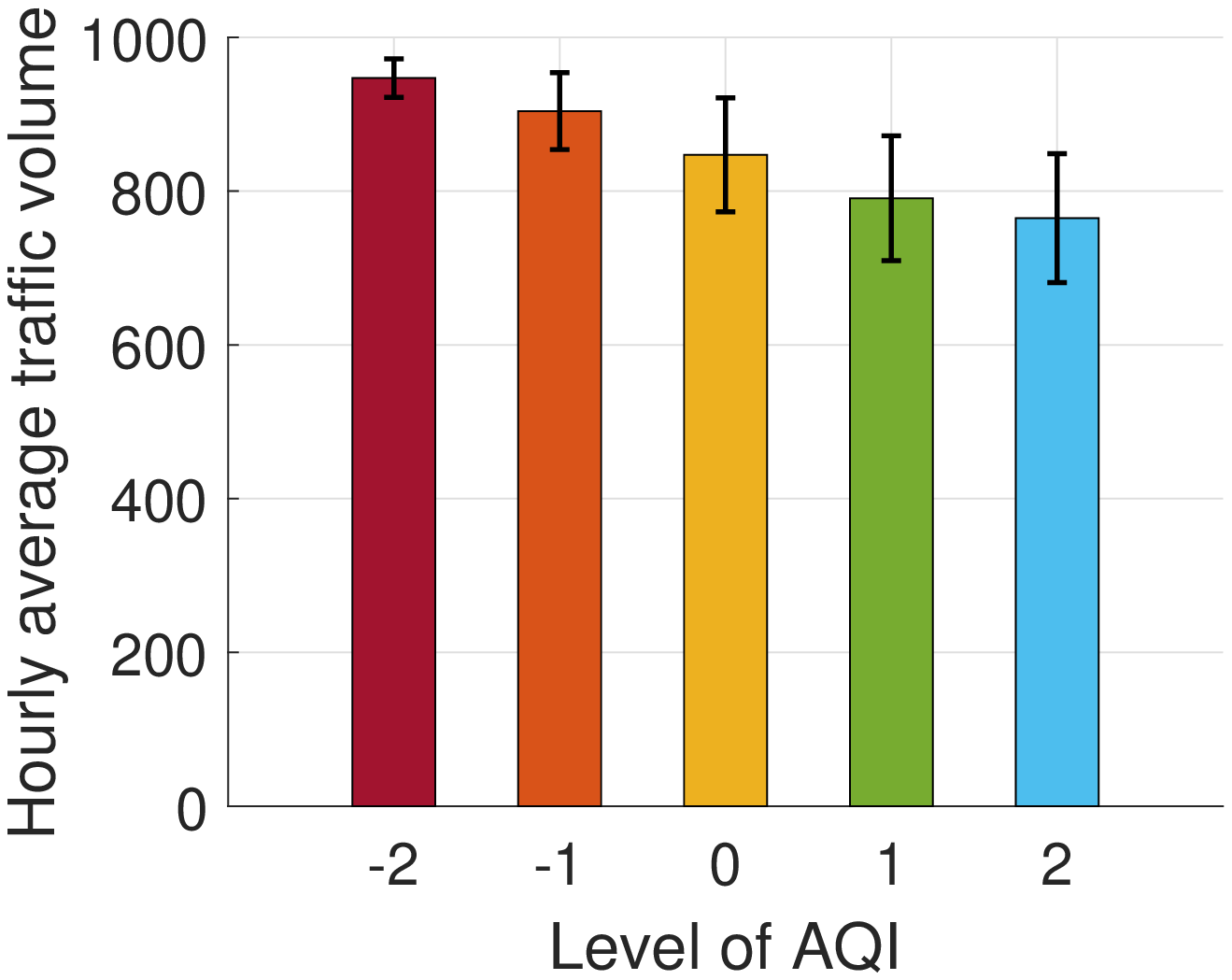}
			\par\vspace{0pt}
		\end{minipage}
	}
\subfloat [Indoor Air Pollutant]
	{
		\label{airpollution_traffic} 
		\begin{minipage}[t]{0.48\linewidth}
			\centering
			\includegraphics[width=0.99\textwidth]{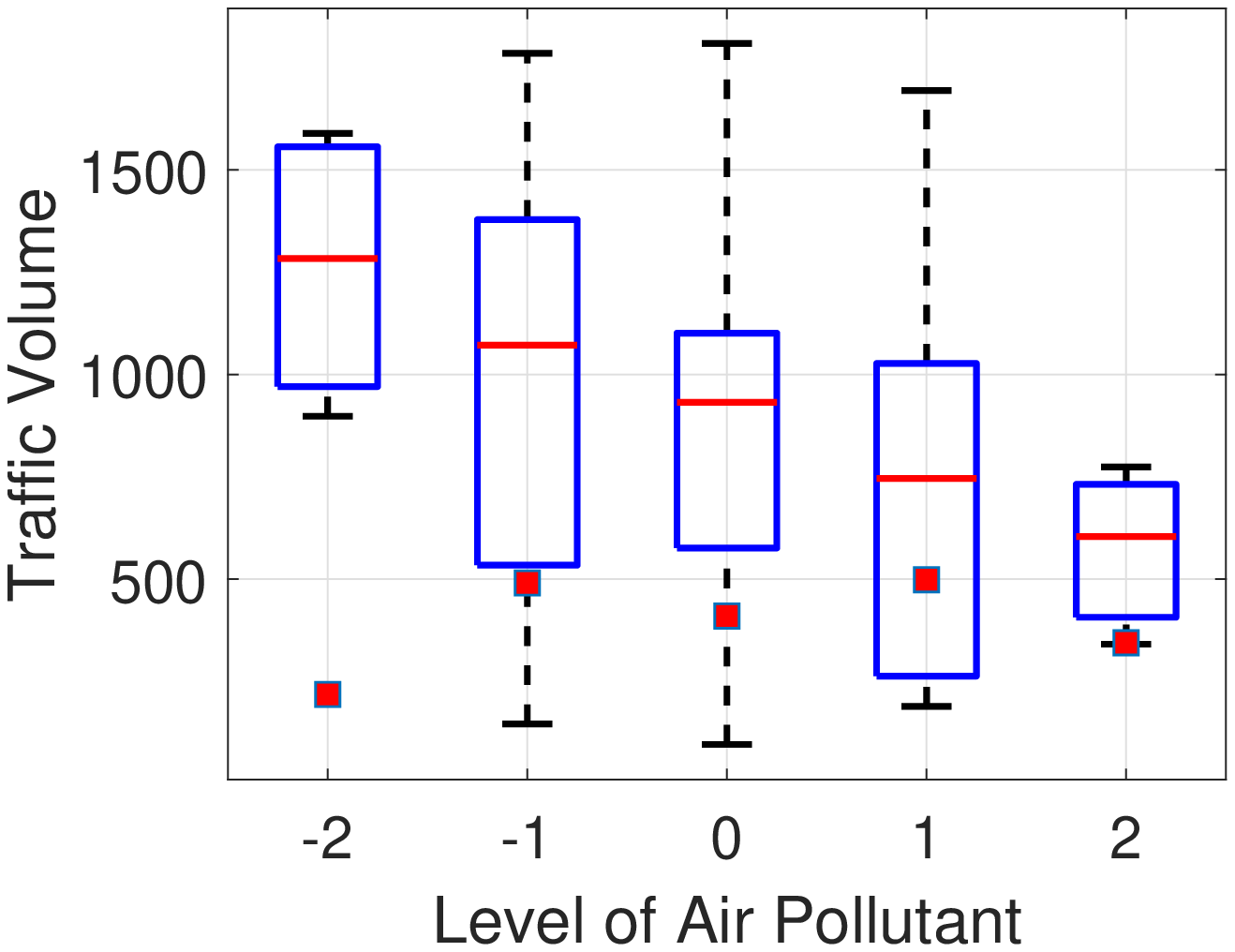}
			\par\vspace{0pt}
		\end{minipage}
	}
	\caption{Correlation analysis between the building environmental data and traffic data (-2 stands for the worst level, 0 for the moderate level and 2 for the best level).}
	\label{environmental_correlations} 

\end{figure}

Besides building occupancy, environmental data have been proved to affect traffic on the nearby roads evidently, and they can also contribute to the improvement of traffic prediction accuracy~\cite{koesdwiady2016improving,lin2018road}.
In this study, for cross-domain traffic prediction,
we typically select 5 types of indoor environmental data (${\rm{C}}{{\rm{O}}_2}$ concentration, building humidity, ${{\rm{O}}_2}$ concentration, building temperature and building air pollution) and 4 types of outdoor environmental data (including outdoor temperature, rainfall, wind speed and air quality index) from the FEIT building.
Through conducting a series of correlation studies with environmental data and traffic volume data,
we find that different types of environmental data have different levels of correlations with traffic volume data.
While some of the environmental data have stronger correlations with outdoor traffic volume,
others show weak correlations.
To this end, we report two representative cases for correlations between environmental data and traffic data as follows.

Fig.~\ref{aqi_traffic} presents correlations between the outdoor Air Quality Index (AQI) and traffic volume by averaging values of traffic data under categorized air quality conditions.
We categorize air quality values into five levels from the worst to the best, represented by -2 to 2.
It can be observed that AQI stays at a lower level when the traffic volume is in higher values.
The main reason is that when the daily traffic volume is at a high level,
more emissions from vehicles can directly pollute the air quality,
thereby affecting outdoor AQI.

Moreover, we evaluate the correlations between traffic volume and indoor air pollutant levels.
As shown in Fig.~\ref{airpollution_traffic},
the box plot illustrates averaged traffic volume with categorized air pollutant levels.
It can be observed that the indoor air pollutant becomes worse with higher traffic volume.
Although indoor environmental factors are not directly related to outdoor traffic volume,
the indoor environmental data can be used as the indirect sensing data for building occupants~\cite{zheng2016urban}.
An essential objective of using indoor environmental data is to assist traffic prediction when the value of building occupancy is at abnormal values.
For example, in case of an emergency evacuation,
there will be a rapid removal of people inside the building,
which will result in higher but false occupancy data generated by smart cameras.
If we combine indoor environmental data in the prediction model,
the prediction errors in the above cases can be eliminated effectively.

To conclude, the environmental data is correlated with the traffic volume;
meanwhile, their correlations are not very strong.
Nevertheless,
the environmental data and the occupancy data can be complementary to each other in cross-domain traffic prediction with building data.

\section{Accurate Traffic Prediction with Cross-Domain Learning of Building Data}\label{architecture}
In this section, based on the correlation analysis in Section~\ref{observation},
we propose a cross-domain learning method for traffic volume prediction by reusing building sensing data.
First, we formulate the traffic volume prediction problem of using cross-domain building sensing data.
Second, to address this problem, we propose a cross-domain learning-based recurrent neural network to learn the non-linear, time-varying, and cross-domain building-traffic correlations for accurate traffic volume prediction.

\subsection{Problem Formulation}
In this work, our goal is to predict traffic volume by
leveraging building sensing data and historical traffic volume data.
Assume that there are $N$ types of building sensing data for $T$ time intervals, we further introduce the following notations.

\begin{itemize}
\item \textbf{Building data types}: a building's IoT sensors generate $N$ types of sensing data that are used for traffic prediction, including ${N_o}$ types of occupancy sensing data (\eg, covering different public zones) and ${N_e}$ types of environmental sensing data (\eg, indoor environmental and outdoor environmental data). Intuitively, $N = {N_o} + {N_e}$.

\item \textbf{Building sensing data}: let $x_t(i)$ denote $i$-th building sensing data at time $t$ and ${{\bf{x}}_t} = \{x_t(i) | 1\leq i \leq N\}$ denote a vector of all building sensing data at time $t$. Accordingly, $\textbf{X}_{[1:T]}={[{\textbf{x}_1}, {\textbf{x}_2},...,{\textbf{x}_T}]} _{T \times n}$ denotes a measurement matrix of all building data across $T$ time intervals.

\item \textbf{Traffic volume data}: let $y_t$ represent traffic volume of a target road segment at time $t$, $1\leq t \leq T-1$. Accordingly, ${\bf{y}}_{[1:T-1]}$ denotes a vector of historical traffic volume of the target road segment across $T-1$ time intervals, where ${\bf{y}} = \{ y(j)|1 \leq j \leq T - 1{\rm{\} }}$.

\item \textbf{Future Traffic volume}: the predicted traffic volume of a target road segment is denoted as $\widehat {\textbf{Y}}_{[T:T+\tau]} =  \{ y(j) | T \leq j \leq T+\tau\}$ while using ${\textbf{Y}}_{[T:T+\tau]}$ to represent its ground truth, where $\tau$ is the time intervals for prediction.
\end{itemize}

\textbf{Problem Definition}: formally,
given $n$ types of building sensing data over $T$ time intervals and historical data of a target road segment over $T-1$ time intervals,
the traffic volume prediction problem for $\tau$ future time intervals is to optimize the prediction errors and defined as:
\begin{eqnarray}\label{equ_Prediction}
&{\rm{\textbf{Minimize}}} &  ||\widehat {\textbf{Y}}_{[T:T+\tau]}-\textbf{Y}_{[T:T+\tau]} ||_F^2, \label{equ_ProblemGoal}\\
&\quad{\rm{where}}   &{\widehat {\bf{Y}}_{[T:T + \tau ]}} = F({{\bf{y}}_{[1:T - 1]}},{{\bf{X}}_{[1:T]}}),
\end{eqnarray}
where $F( \cdot )$ denotes the nonlinear mapping function from building sensing data to traffic data that a predicting model needs to learn.


\subsection{Attention mechanisms-based encoder-decoder Recurrent Neural Network}
\label{Encoder-method}
\begin{figure}[t]
	\centering
	\includegraphics[width=0.5\textwidth]{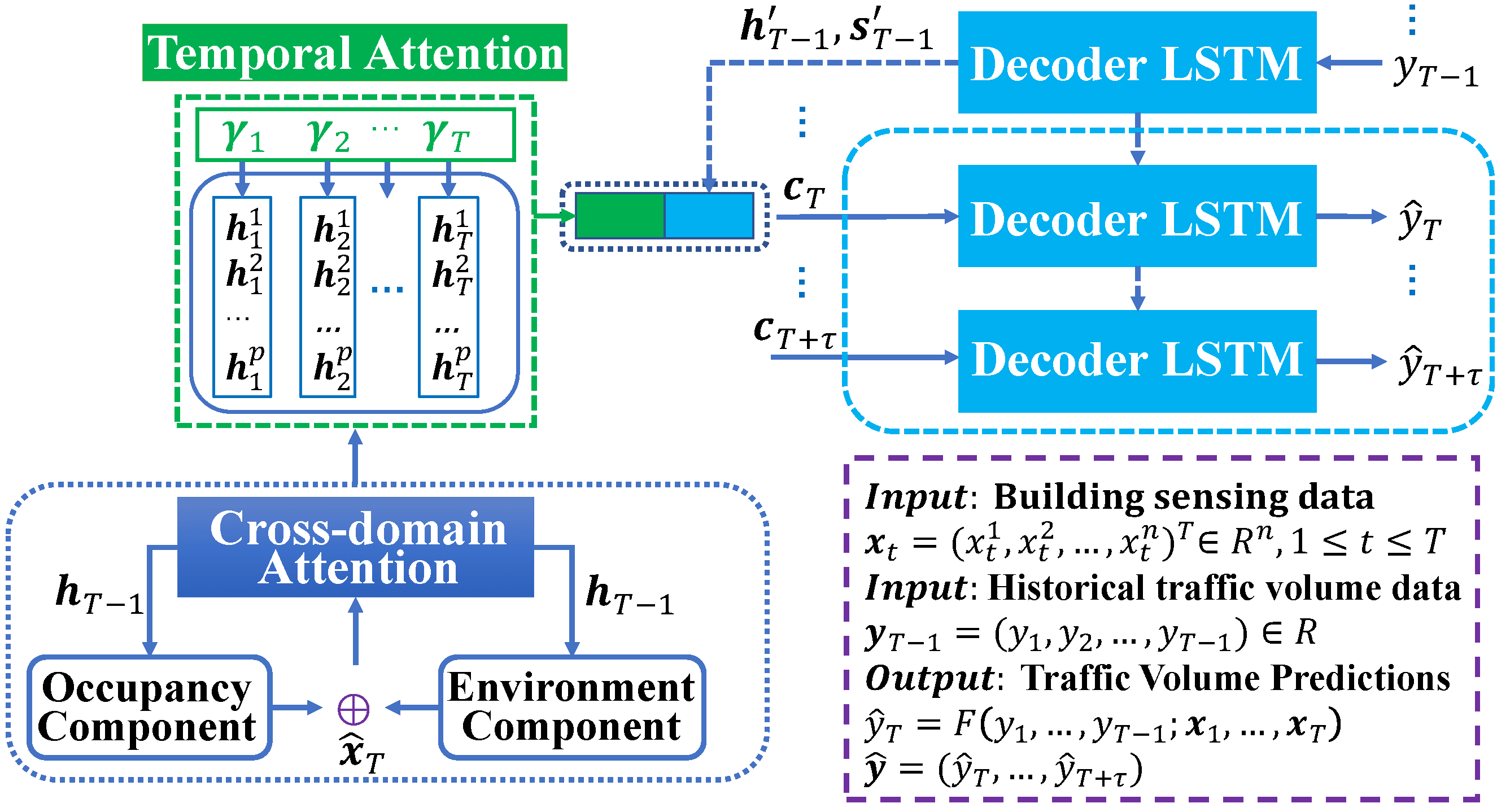}
	\caption{The graphical architecture of cross-domain attention-based recurrent neural networks for cross-domain traffic prediction.}
	\label{graphical_architecture}
\end{figure}

To solve the cross-domain learning-based traffic volume prediction problem,
we propose \BuildSenSys, an LSTM based encoder-decoder architecture with dual-attentions mechanisms.
First, we employ cross-domain attention on input data to capture building-traffic correlations and adaptively select the most relevant building sensing data at each step of prediction.
Second, we apply temporal attention to capture temporal features from historical dependencies.
Then, \BuildSenSys adaptively select the most relevant encoder hidden states across all time intervals.
By integrating the above attention mechanisms with LSTM-based recurrent neural networks,
we can further jointly the train \BuildSenSys model with standard back propagation.
As a result, \BuildSenSys is capable of selecting the most relevant building sensing data for traffic prediction and capture long-term temporal features of traffic volume.
The overall framework of \BuildSenSys is presented in Fig.~\ref{graphical_architecture}.

\subsubsection{Encoder with cross-domain attention}
The encoder in the \BuildSenSys framework is an LSTM-based recurrent neural network,
and it encodes the input sequence into a feature vector.
For cross-domain traffic volume prediction,
given $N$ types of building sensing data,
we denote the input sequence as $\textbf{X}={({\textbf{x}_1}, ...,{\textbf{x}_t},...,{\textbf{x}_T})}$, where ${{\bf{x}}_t} \in {\mathbb{R}^N}$.
The hidden state of the encoder at time interval $t$ is computed by
\begin{equation}\label{initial_lstm}
{{\bf{h}}_t} = {f_e}({{\bf{h}}_{t - 1}},{{\bf{x}}_t}),
\end{equation}
where ${{\bf{h}}_{t - 1}} \in {\mathbb{R}^p}$ is the previous hidden state of the encoder at time interval $t-1$;
$p$ is the size of the hidden state in the encoder; and $f_e$ is an LSTM based recurrent neural network.
Since LSTMs are capable of learning long-term dependencies, we employ the classic LSTM unit that is with one memory cell and three sigmoid gates.
At time interval $t$, the cell state of memory is ${{\bf{s}}_t}$, the forget gate is ${{\bf{f}}_t}$, and the input gate is ${{\bf{i}}_t}$.
The encoder LSTM updates its hidden state by
\begin{equation}\label{lstm_1}
{{\bf{f}}_t} = \sigma ({{\bf{W}}_f}[{{\bf{h}}_{t - 1}};{{\bf{x}}_t}] + {{\bf{b}}_f}),
\end{equation}
\begin{equation}\label{lstm_2}
{{\bf{i}}_t} = \sigma ({{\bf{W}}_i}[{{\bf{h}}_{t - 1}};{{\bf{x}}_t}] + {{\bf{b}}_i}),
\end{equation}
\begin{equation}\label{lstm_3}
{{\bf{o}}_t} = \sigma ({{\bf{W}}_o}[{{\bf{h}}_{t - 1}};{{\bf{x}}_t}] + {{\bf{b}}_o}),
\end{equation}
\begin{equation}\label{lstm_4}
{{\bf{s}}_t} = {{\bf{f}}_t} \odot {{\bf{s}}_{t - 1}} + {{\bf{i}}_t} \odot tanh({{\bf{W}}_s}[{{\bf{h}}_{t - 1}};{{\bf{x}}_t}] + {{\bf{b}}_s}),
\end{equation}
\begin{equation}\label{lstm_5}
\textbf{h}_t = \textbf{o}_t\odot tanh(\textbf{s}_{t}),
\end{equation}
where $[ \cdot ; \cdot ]$ is a concatenation operation; $\sigma$ is a logistic sigmoid function; $\odot$ is a pointwise multiplication;
${{\bf{W}}_f}$, ${{\bf{W}}_i}$, ${{\bf{W}}_o}$, ${{\bf{W}}_s}$ and
${{\bf{b}}_f}$, ${{\bf{b}}_i}$, ${{\bf{b}}_o}$, ${{\bf{b}}_s}$ are the learnable parameters.

Inspired by that visual attention allows human to focus on a certain region of
images or sentences for creating the perception of information,
attention mechanism has become an integral part of the compelling sequence modeling and
transduction models~\cite{vaswani2017attention}.
In general, attention mechanisms are proposed to make a soft selection over historical data
by calculating and assigning different weights to them.
In this work, we aim to predict the traffic volume with different types of building sensing data and historical traffic volume data.
To achieve this goal, we propose cross-domain attention to capture complex building-traffic correlations
and enhance feature representation of all input data.
In specific, based on the results of cross-domain correlation analysis,
we propose to learn different correlations with the occupancy component and environmental component, respectively.

\textbf{Occupancy Component:} we envision a multi-zone scenario for traffic volume prediction,
with ${j_{th}}$ zone's occupancy as ${{\bf{x}}^j} = {(x_1^j,x_2^j,...,x_t^j)^ \top } \in {\mathbb{R}^T}$ and $1 \le j \le {N_o}$.
The cross-domain attention on occupancy components is calculated by referring the previous hidden state ${{\bf{h}}_{t - 1}}$ and previous cell state ${{\bf{s}}_{t - 1}}$ of encoder LSTM by
\begin{equation}
o_t^j = {\bf{v}}_o^ \top \tanh ({{\bf{W}}_o}[{{\bf{h}}_{t - 1}};{{\bf{s}}_{t - 1}}] + {{\bf{U}}_o}{{\bf{x}}^j} + {{\bf{b}}_o}),
\end{equation}
\begin{equation}
\beta _t^j = \frac{{\exp (o_t^j)}}{{\sum\nolimits_{i = 1}^{{N_o}} {\exp (o_t^i)} }},
\end{equation}
where ${{\bf{v}}_o}, {{\bf{b}}_o} \in {\mathbb{R}^T}$,
${{\bf{W}}_o} \in {\mathbb{R}^{T \times 2p}}$ and
${{\bf{U}}_o} \in {\mathbb{R}^{T \times T}}$ are learning parameters.
By using another softmax function,
we ensure that all attention weights with occupancy component are normalized, and the total weight for all occupancy input data is 1.
With cross-domain attention weights for all public zones,
we acquire the output vector from the occupancy component at time interval $t$ as:
\begin{equation}
{\bf{\hat x}}_t^{occ} = {(\beta _t^1x_t^1,...\beta _t^jx_t^j,...\beta _t^{{N_o}}x_t^{{N_o}})^ \top }.
\end{equation}

\textbf{Environmental Component:}
given ${k_{th}}$ type of environmental data input as ${{\bf{x}}^k}$,
where ${{\bf{x}}^k} = {(x_1^k,x_2^k,...,x_t^k)^ \top } \in {\mathbb{R}^T}$ and $1 \le k \le {N_e}$.
We employ the cross-domain attention for the environmental component to adaptively capture the dynamic correlation between traffic volume and ${k_{th}}$ type of environmental data by
\begin{equation}
e_t^k = {\bf{v}}_e^ \top \tanh ({{\bf{W}}_e}[{{\bf{h}}_{t - 1}};{{\bf{s}}_{t - 1}}] + {{\bf{U}}_e}{{\bf{x}}^k} + {{\bf{b}}_e}),
\end{equation}
\begin{equation}
\alpha _t^k = \frac{{\exp (e_t^k)}}{{\sum\nolimits_{i = 1}^{{N_e}} {\exp (e_t^i)} }},
\end{equation}
where $[ \cdot ; \cdot ]$ is a concatenation operation.
Here, ${{\bf{v}}_e}, {{\bf{b}}_e} \in {\mathbb{R}^T}$,
${{\bf{W}}_e} \in {\mathbb{R}^{T \times 2p}}$ and
${{\bf{U}}_e} \in {\mathbb{R}^{T \times T}}$ are learnable parameters.
By applying a softmax function to $e_t^k$,
we obtain the normalized attention weight $\alpha _t^k$ for ${k_{th}}$ environmental data at the time interval $t$.
With cross-domain attention weights for all types of environmental data input,
the output vector of the environmental component at time interval $t$ can be adaptively acquired by
\begin{equation}
{\bf{\hat x}}_t^{env} = {(\alpha _t^1x_t^1,...\alpha _t^kx_t^k,...\alpha _t^{{N_e}}x_t^{{N_e}})^ \top }.
\end{equation}

Finally, for encoder LSTM, we adaptively concatenate above output vectors from different components and extract the final output vector of cross-domain attention mechanism as:
\begin{equation}
{{{\bf{\hat x}}}_t} = [{\bf{\hat x}}_t^{env};{\bf{\hat x}}_t^{occ}],
\end{equation}
where ${{\hat x}_t} \in {\mathbb{R}^N}$.
We feed the final output vector ${{{\bf{\hat x}}}_t}$ into the encoder LSTM as its new input at time interval $t$. Consequently,
the hidden state of encoder LSTM in Eq.~\ref{initial_lstm} is updated by
\begin{equation}
{{\bf{h}}_t} = {f_e}({{\bf{h}}_{t - 1}},{{{\bf{\hat x}}}_t}),
\end{equation}
where $f_e$ is the encoder LSTM network described in Eq.~\ref{lstm_1} to Eq.~\ref{lstm_5}.

\subsubsection{Decoder with temporal attention}
\label{Decoder-method}
The basic idea of attention mechanism is to distinguish task-related importance of input data in historical time intervals.
Intuitively, the decoder with a temporal attention mechanism can be trained to capture temporal dependencies between traffic volume and building data by calculating attention weights for all encoder hidden states.
With the temporal attention weights assigned to all encoder hidden states,
decoder LSTM can focus on the most relevant input data when decoding~\cite{vaswani2017attention}.
To this end, we employ a temporal attention mechanism that enables the decoder to select relevant encoder hidden states across all time intervals adaptively.
In specific, when computing the attention vector for encoder hidden state at time interval $t$, the temporal attention mechanism refers to the previous hidden state ${{{\bf{h'}}}_{t - 1}}$ and previous cell state ${{{\bf{s'}}}_{t - 1}}$ of decoder LSTM by
\begin{equation}
\mu _t^i = {\bf{v}}_d^ \top \tanh ({{\bf{W}}_d}[{{{\bf{h'}}}_{t - 1}};{{{\bf{s'}}}_{t - 1}}] + {{\bf{U}}_d}{{\bf{h}}_i} + {{\bf{b}}_d}),1 \le i \le T,
\end{equation}
\begin{equation}
\gamma _t^i = \frac{{\exp (\mu _t^i)}}{{\sum\nolimits_{l = 1}^T {\exp (\mu _t^l)} }},
\end{equation}
where $[\cdot]$ is a concatenation operation;
${{\bf{v}}_d},{{\bf{b}}_d} \in {\mathbb{R}^p}$;
${{\bf{W}}_d} \in {\mathbb{R}^{p \times 2q}}$;
${{\bf{U}}_d}\in {\mathbb{R}^{p \times p}}$ are learning parameters.
Through a softmax layer,
the temporal attention weight $\gamma _t^i$ is calculated for ${i_{th}}$ encoder hidden state at time interval $t$.
As the component of the input sequence is temporally mapped to each encoder,
we calculate the context vector ${{\bf{c}}_t}$ as a weighted sum of all encoder hidden states by
\begin{equation}
{{\bf{c}}_t} = \sum\limits_{i = 1}^T {\gamma _t^i{{\bf{h}}_i}}.
\end{equation}
To capture the dynamic temporal correlation in traffic volume data,
we further combine the context vector with ${\bf{y}} = ({y_1},...,{y_t},...,{y_{T - 1}})$ as follows:
\begin{equation}
{{\tilde y}_{t - 1}} = {{\tilde w}^ \top }[{y_{t - 1;}}{{\bf{c}}_{t - 1}}] + \tilde b,
\end{equation}
\begin{equation}
{{{\bf{h'}}}_t} = {f_d}({{{\bf{h'}}}_{t - 1}},{{\tilde y}_{t - 1}}),
\end{equation}
where ${f_d}$ is an LSTM-based recurrent neural network as decoder;
${\tilde w} \in {\mathbb{R}^{p+1}}$ and ${\tilde b} \in \mathbb{R}$ are parameters to map the concatenation result to the size of the decoder input.
As the structure of LSTM unit in the decoder is exactly the same as the encoder
(referred to Eq.~\ref{lstm_1} to \ref{lstm_5}), we omit the update process of ${f_d}$.
Finally, attention mechanisms based recurrent neural network concatenates the context vector ${{\bf{c}}_T}$ with decoder hidden state ${{{\bf{h'}}}_T}$,
predicting the traffic volume at time interval $T$ as
\begin{equation}
{{\hat y}_T} = {\bf{v}}_y^ \top ({{\bf{W}}_y}[{{\bf{c}}_T};{{{\bf{h'}}}_T}] + {{\bf{b}}_y}) + b,
\end{equation}
where $[{{\bf{c}}_T};{{{\bf{h'}}}_T}] \in {\mathbb{R}^{p+q}}$ is a concatenation operation;
parameters ${{\bf{W}}_y} \in {\mathbb{R}^{q \times (p + q)}}$ and ${{\bf{b}}_y} \in {\mathbb{R}^{q}}$ together map the concatenation to the size of decoder hidden states.
The final output is generated by a mapping function with weights
${{\bf{v}}_y} \in {\mathbb{R}^{q}}$ and ${{\bf{b}}_y} \in \mathbb{R}$.

\begin{figure}[t]
	\centering
	\includegraphics[width=0.5\textwidth]{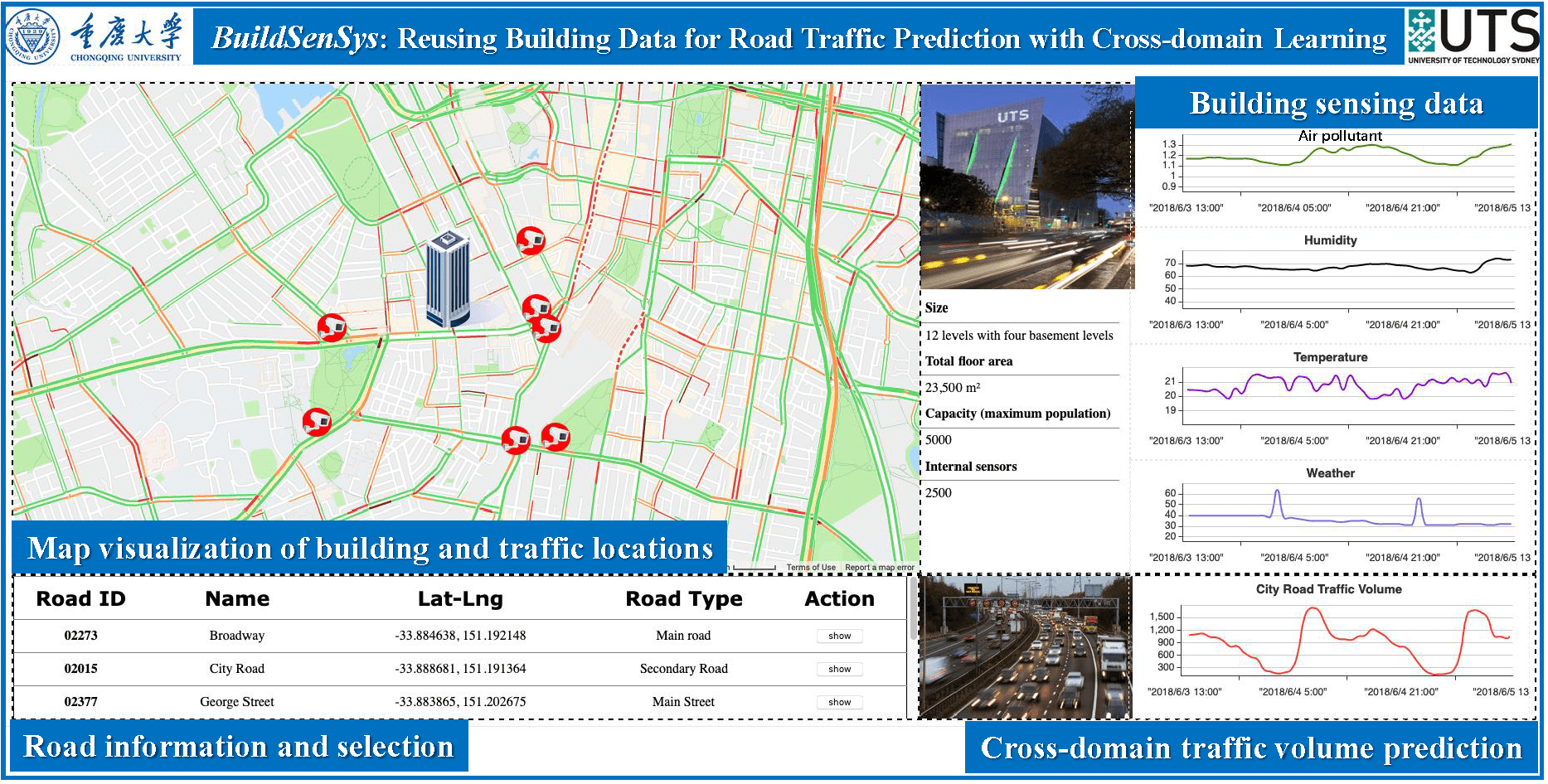}
	\caption{The prototype system of \BuildSenSys for data visualization and traffic prediction.}
	\vspace{0.1in}
	\label{case_study}
\end{figure}

\section{Performance Evaluation}\label{simulation}
To evaluate the performance of \BuildSenSys, we develop a prototype system,
as shown in Fig.~\ref{case_study}.
We first train the cross-domain learning-based RNN model
with a training set of building sensing data and traffic data.
With the pre-trained model, \BuildSenSys can output predicted traffic volume on nearby roads
only with building sensing data as the input.
In the following, we first introduce the experimental settings,
including dataset, baseline methods, evaluation metrics, and model parameters.
Then, we conduct extensive experimental studies on \BuildSenSys and evaluate its performance in terms of baseline comparison, parameter study, ablation study, attention weight, and other extensive comparisons.

\begin{figure*}[t]
	\centering
	\subfloat [Road A]
	{
		\label{cityroad}
		\begin{minipage}[t]{0.24\linewidth}
			\centering
			\includegraphics[width=0.99\textwidth]{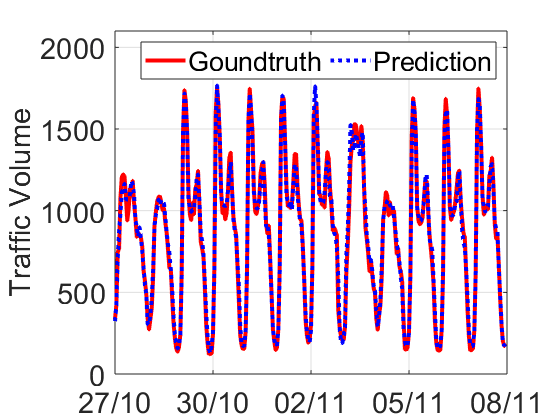}
			\par\vspace{0pt}
		\end{minipage}
	}
    \subfloat [Road B]
	{
		\label{paramatta}
		\begin{minipage}[t]{0.24\linewidth}
			\centering
			\includegraphics[width=0.99\textwidth]{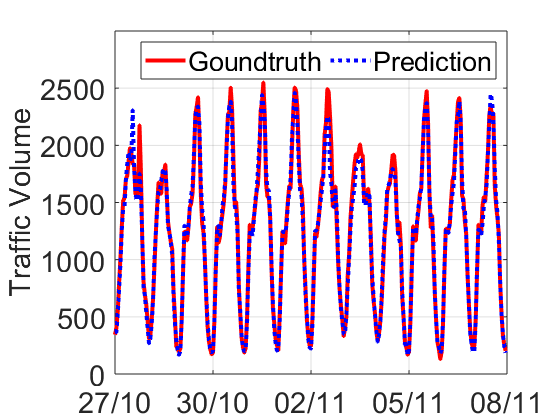}
			\par\vspace{0pt}
		\end{minipage}
	}
    \subfloat [Road C] 
	{
		\label{georgestreet}
		\begin{minipage}[t]{0.24\linewidth}
			\centering
			\includegraphics[width=0.99\textwidth]{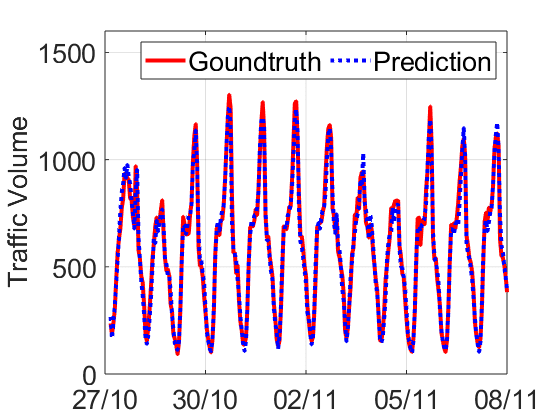}
			\par\vspace{0pt}
		\end{minipage}
	}
	\subfloat [Road D]
	{
		\label{regentstreet}
		\begin{minipage}[t]{0.24\linewidth}
			\centering
			\includegraphics[width=0.99\textwidth]{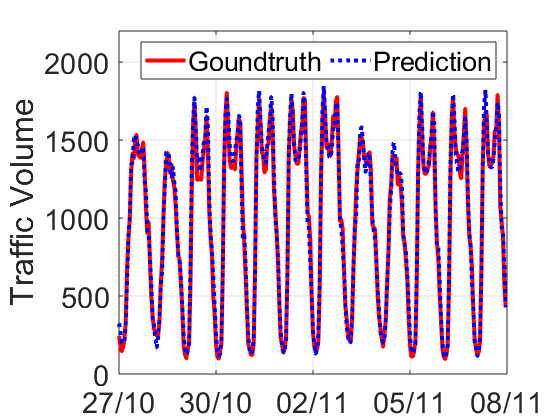}
			\par\vspace{0pt}
		\end{minipage}
	}
	\caption{Comparison between predicted traffic volume of \BuildSenSys and the ground truth on four different roads.}
	\label{traffic_prediction}
\end{figure*}

\subsection{Experimental Methodology and Settings} \label{Settings}

\subsubsection{Dataset Description}
Based on the results of the observation study,
we select traffic data from nearby roads and relevant building sensing data as the training input data for the \BuildSenSys model.
For traffic data, we collect traffic volume count data from the official website of
the Department of Roads and Maritime Services, New South Wales State~\cite{nsw2018traffic}.
The traffic volume count data is generated by permanent and temporary roadside collection devices,
monitoring the number of passing vehicles on each road with the one-hour interval.
We collect 12-month traffic volume data (from 2018/1/1 to 2018/12/31)
on four nearby roads of the building.
For building sensing data,
we collect data from three categories,
\ie, building occupancy, building environmental data, outdoor environmental data.
First, building occupancy data is generated by camera sensors distributed at Point-of-Interest Zones inside the building.
We process and aggregate these data with the PLCount algorithm~\cite{sangoboye2016plcount} for overall building occupancy.
Second, the building environmental data include ${\rm{C}}{{\rm{O}}_2}$ concentration, building humidity, ${{\rm{O}}_2}$ concentration, building temperature, and building air pollution.
Third, the outdoor environmental data is collected from the rooftop weather station of our building,
including outdoor temperature, rainfall, and wind speed.
Besides, as vehicle emissions have a direct influence on outdoor air quality,
we further adopt the hourly AQI data from the Bureau of Meteorology's official website and
integrate the AQI data into outdoor environmental data for traffic prediction.
At last, we synchronize all building sensing data and traffic volume data with the one-hour interval for training and testing purposes.

\subsubsection{Baseline Methods}
To comprehensively evaluate the performance of \BuildSenSys, we compare it with seven baseline methods as follows.
\begin{itemize}
  \item \textbf{HA}~\cite{zhang2017deep}: The historical average (HA) model, which predicts the traffic volume by averaging the historical value of all corresponding time intervals.
  \item \textbf{ARIMA}~\cite{min2011real}: The autoregressive integrated moving average (ARIMA) model, which is a classic model to predict future time series.
  \item \textbf{VAR}~\cite{chandra2009predictions}: The Vector Auto-regressive (VAR) model is an extension of the univariate autoregressive model, which has been widely used for multivariate time series forecasting.
  \item \textbf{LWR}~\cite{zheng2016urban}: The Locally Weighted Linear Regression (LWR) model is a non-parametric model, which performs regressions around points of interest.
  \item \textbf{LSTM}~\cite{zhao2017lstm}: The Long Short-Term Memory (LSTM) network is a variation of recurrent neural networks designed for avoiding the vanishing gradient problem.
  \item \textbf{Seq2Seq}~\cite{sutskever2014sequence}: The Sequence to Sequence model based on an encoder-decoder architecture and recurrent neural networks, consisting of three parts, \ie, encoder, context vector, and decoder.
  \item \textbf{Seq2Seqw/attn}~\cite{sutskever2014sequence}: The Sequence to Sequence model with a temporal attention mechanism.
\end{itemize}
Note that all baseline methods take the time of a day as an essential feature of input data, and they use this feature for traffic prediction from different perspectives.
For example, the HA leverages time of the day with statistical regression, while ARIMA and VAR process this feature with autoregression. Moreover, the neural networks of the LSTM and Seq2Seq models learn the time of the day as a feature with their hidden states.
In the proposed \BuildSenSys model, we further address the time of the day by employing a temporal attention mechanism to learn temporal features of traffic data.

\subsubsection{Evaluation Metrics and Parameter Settings}
To evaluate the prediction accuracy, we employ three widely used evaluation metrics, \ie,
Mean Absolute Error (MAE), Root Mean Squared Error (RMSE) and Mean Absolute Percentage Error (MAPE)~\cite{qin2017dual}.
Both MAE and RMSE are scale-dependent metrics, and MAPE is a scale-independent metric.
Specifically, MAE measures the average magnitude of errors in prediction results as Eq.~\ref{Equ_MAE}.
RMSE measures square root of the average squared differences between prediction results and ground truth as Eq.~\ref{Equ_RMSE}.
MAPE measures the size of errors in percentage terms to quantify the prediction accuracy as Eq.~\ref{Equ_MAPE}.
\begin{equation}\label{Equ_MAE}
{\bf{MAE}} = \frac{1}{\tau }\sum\nolimits_{[T:T{\rm{ + }}\tau ] \in {\mathbb{R}^T}} {\left| {{{\bf{Y}}_{[T:T{\rm{ + }}\tau ]}} - {{{\bf{\hat Y}}}_{[T:T{\rm{ + }}\tau ]}}} \right|},
\end{equation}

\begin{equation}\label{Equ_RMSE}
{\bf{RMSE}} = \sqrt {\frac{1}{\tau }\sum\nolimits_{[T:T{\rm{ + }}\tau ] \in {\mathbb{R}^T}} {{{({{\bf{Y}}_{[T:T{\rm{ + }}\tau ]}} - {{{\bf{\hat Y}}}_{[T:T{\rm{ + }}\tau ]}})}^{\rm{2}}}} },
\end{equation}

\begin{equation}\label{Equ_MAPE}
{\bf{MAPE}} = \frac{1}{\tau }\sum\nolimits_{[T:T{\rm{ + }}\tau ] \in {\mathbb{R}^T}} {\left| {\frac{{{{\bf{Y}}_{[T:T{\rm{ + }}\tau ]}} - {{{\bf{\hat Y}}}_{[T:T{\rm{ + }}\tau ]}}}}{{{{\bf{Y}}_{[T:T{\rm{ + }}\tau ]}}}}} \right|},
\end{equation}
where ${{{{\bf{\hat Y}}}_{[T:T{\rm{ + }}\tau ]}}}$ and ${{{\bf{Y}}_{[T:T{\rm{ + }}\tau ]}}}$ are the prediction results of traffic volume from time interval $T$ to $T{\rm{ + }}\tau$, respectively.


In this work, the proposed \BuildSenSys model is implemented with Tensorflow framework and trained together with other baseline models on two NVIDIA Quadro P5000 GPUs with 16 GB memory.
For model training,
the tunable hyperparameters for \BuildSenSys include
the time window $L$ (\ie, the length of input data in hours),
the length of predicting window $\tau$ (the number of days for future traffic prediction) and the size of hidden states in encoder/decoder~(denoted by ${h_a}$ and ${h_b}$, respectively).
For LSTM, Seq2Seq, Seq2Seqw/attn, and \BuildSenSys model,
${h_a}$ and ${h_b}$ are tuned from 32, 64, 126, 256 to 512, and $L$ is tuned from 4, 6, 12, 18, 24 to 48 (hours), respectively.
In the training session, the batch size, learning rate, and dropout rate are set to 256, 0.001, and 0.2, respectively, and the Adam is the optimizer for the \BuildSenSys model.
For all building sensing data and traffic volume data,
we split the dataset into the training set (70\%), validation set (10\%) and test set (20\%) in chronological order.

\begin{table*}[t]
\caption{Performance comparison with baseline methods on different roads.}
\centering
\begin{tabular}{c|c|c|c|c|c|c|c|c|c|c|c|c}
\hline
\multirow{2}{*}{Models} & \multicolumn{3}{c|}{Road A} & \multicolumn{3}{c|}{Road B} & \multicolumn{3}{c|}{Road C} & \multicolumn{3}{c}{Road D} \\ \cline{2-13}
                        & RMSE     & MAE    & MAPE               & RMSE      & MAE     &MAPE      &RMSE       & MAE     & MAPE    & RMSE     & MAE    & MAPE    \\ \hline\hline
LWR                     & 244.51   & 155.69    & 22.78\%         & 257.86    & 176.94  &25.91\%   &206.52     &176.35   &22.61\%   & 239.48   & 190.79   & 22.13\%     \\ \hline
ARIMA                   & 152.74   & 141.22    & 14.51\%         & 192.73    & 142.66  &15.58\%   &149.11     &138.09   &14.70     & 173.32 & 124.31   & 14.22\%       \\ \hline
VAR                     & 117.35   & 110.83    & 12.99\%         & 120.37    & 116.9   &13.06\%   &98.65      &94.15    &11.05     & 124.61   & 110.68   & 12.46\%       \\ \hline
HA                      & 108.01   & 89.72     & 11.37\%         & 125.88    & 93.57   &12.78\%   &90.71      &65.20    &10.26\%   & 117.03   & 89.45    & 11.12\%       \\ \hline
LSTM                    & 74.37    & 58.30     & 9.83\%          & 91.71     & 72.47   &11.04\%   &70.71      &58.9     &9.31\%    & 90.18    & 67.37    & 10.39\%       \\ \hline
Seq2Seq                 & 72.86    & 51.06     & 7.40\%          & 89.25     & 64.78   &8.29\%    &66.19      &53.32    &7.12\%    & 87.22    & 63.20    & 8.14\%        \\ \hline
Seq2Seqw/attn (128)     & 59.58    & 45.18     & 6.89\%          & 73.5      & 59.12   &7.75\%    &54.07      &46.98    &6.73\%    & 81.15    & 56.95    & 7.49\%        \\ \hline
Seq2Seqw/attn (512)     & 50.18    & 39.74     & 5.71\%          & 68.71     & 50.71   &6.97\%    &45.51      &37.33    &5.49\%    & 65.3     & 46.72    & 6.4\%         \\ \hline
\BuildSenSys (128)     & 46.89    & 33.35     & 4.85\%          & 60.18     & 41.89   & 4.36\%   &36.42      &27.14    &5.25\%    & 50.45    & 38.56    & 5.2\%         \\ \hline
\bf{\BuildSenSys (512)}     & \bf{33.08} & \bf{22.78}  & \bf{2.05}\% &\bf{58.33} & \bf{39.71} &\bf{3.64}\%       &\bf{30.49} &\bf{20.29} &\bf{2.51}\%   & \bf{35.10} & \bf{25.75} & \bf{2.22}\%   \\ \hline
\end{tabular}
\label{baseline_table}
\end{table*}

\subsection{Experimental Evaluations}
\subsubsection{Evaluations on overall prediction results}

First, we evaluate the overall prediction accuracy of \BuildSenSys by comparing prediction results with the ground truth on four different types of roads, \ie,
roads A (minor highway), B (major highway), C (main street), and D (primary street).
Fig.~\ref{traffic_prediction} illustrates the predicted traffic volume and the ground truth over 12 days.
As shown in Fig.~\ref{cityroad} to Fig.~\ref{regentstreet}, the prediction results are highly close to the ground truth.
Thus, \BuildSenSys can successfully capture the cross-domain correlations and temporal dependencies in reusing building data for traffic volume prediction.
Moreover, the above results indicate that the distance between a road segment and the building can significantly affect the prediction accuracy.
For example, as illustrated in Fig.~\ref{paramatta},
prediction errors on road B are much larger than those of roads A, C and D.
As shown in Table~\ref{ComparisonResults}, road B is the furthest (3.0 km) to the building among all four roads.

\subsubsection{Comparisons with Baselines}
\begin{figure*}[t]
	\centering
	\subfloat [RMSE]
	{
		\label{hidden_comparison_rmse}
		\begin{minipage}[t]{0.3\linewidth}
			\centering
			\includegraphics[width=0.99\textwidth]{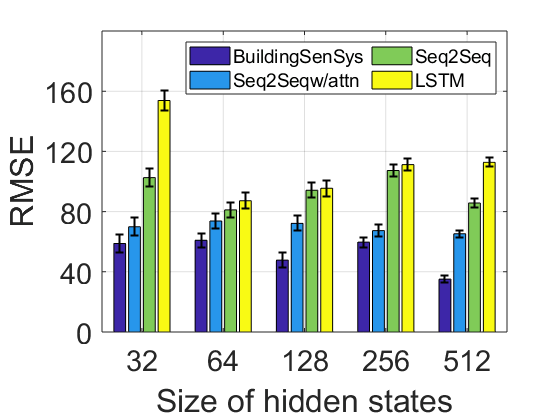}
			\par\vspace{0pt}
		\end{minipage}
	}
	\subfloat [MAE]
	{
		\label{hidden_comparison_mae}
		\begin{minipage}[t]{0.3\linewidth}
			\centering
			\includegraphics[width=0.99\textwidth]{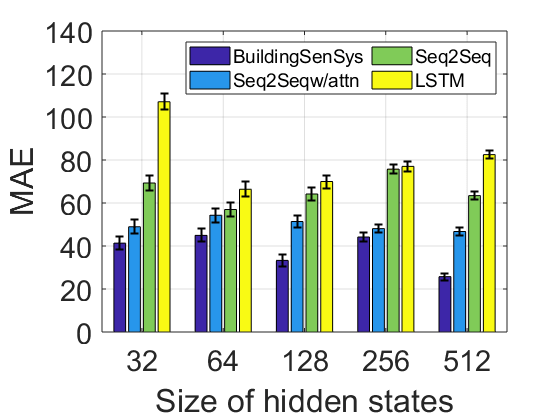}
			\par\vspace{0pt}
		\end{minipage}
	}
    \subfloat [MAPE]
	{
		\label{hidden_comparison_mape}
		\begin{minipage}[t]{0.3\linewidth}
			\centering
			\includegraphics[width=0.99\textwidth]{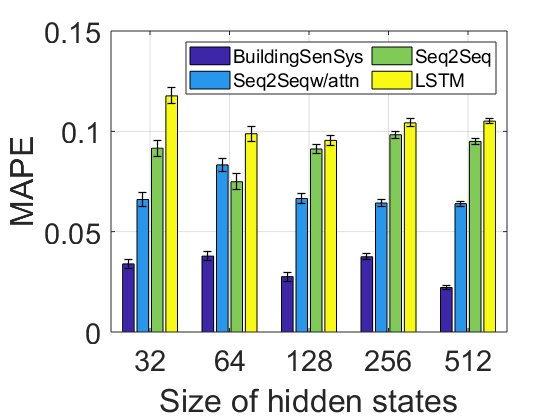}
			\par\vspace{0pt}
		\end{minipage}
	}
	\caption{Impact of the hidden states on the performance of \BuildSenSys and three RNN-based baseline methods.}
	\label{hidden_comparison}
\end{figure*}

\begin{figure*}[t]
	\centering
	\subfloat [RMSE]
	{
		\label{timewindow_comparison_mae}
		\begin{minipage}[t]{0.3\linewidth}
			\centering
			\includegraphics[width=0.99\textwidth]{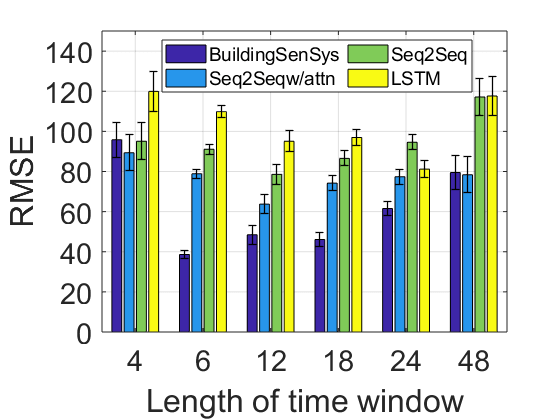}
			\par\vspace{0pt}
		\end{minipage}
	}
	\subfloat [MAE]
	{
		\label{timewindow_comparison_rmse}
		\begin{minipage}[t]{0.3\linewidth}
			\centering
			\includegraphics[width=0.99\textwidth]{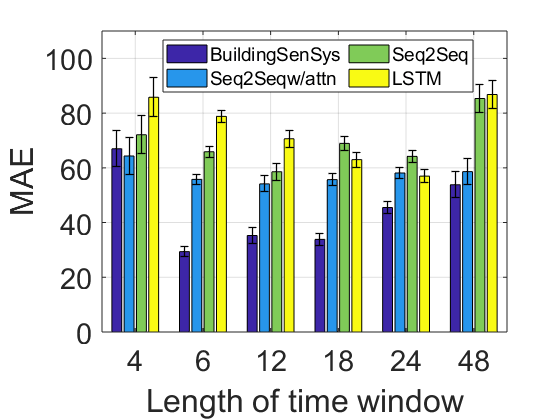}
			\par\vspace{0pt}
		\end{minipage}
	}
    \subfloat [MAPE]
	{
		\label{timewindow_comparison_mape}
		\begin{minipage}[t]{0.3\linewidth}
			\centering
			\includegraphics[width=0.99\textwidth]{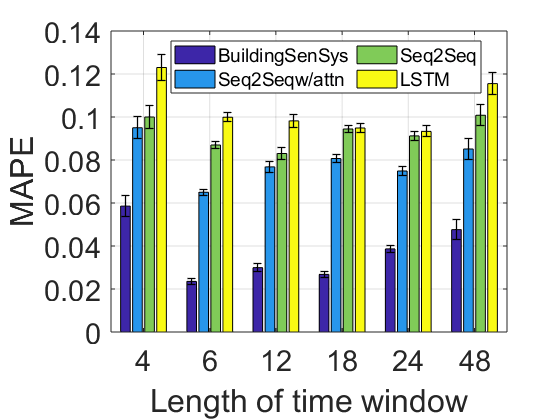}
			\par\vspace{0pt}
		\end{minipage}
	}
	\caption{Impact of the input time window on the performance of \BuildSenSys and three RNN-based baseline methods.}
	\label{timewindow_comparison}
\end{figure*}

\begin{figure*}[t]
	\centering
	\subfloat [Root Mean Squared Error]
	{
		\label{future_metrics_rmse}
		\begin{minipage}[t]{0.3\linewidth}
			\centering
			\includegraphics[width=0.99\textwidth]{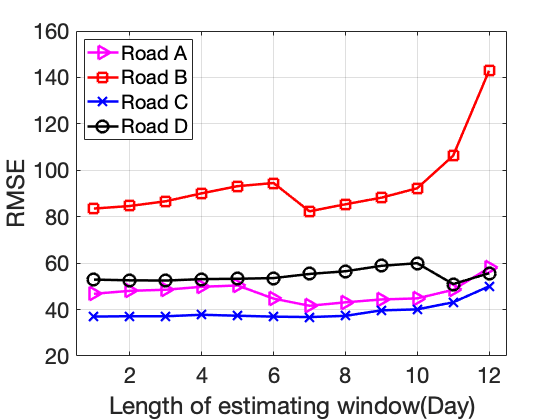}
			\par\vspace{0pt}
		\end{minipage}
	}
	\subfloat [Mean Average Error]
	{
		\label{future_metrics_mae}
		\begin{minipage}[t]{0.3\linewidth}
			\centering
			\includegraphics[width=0.99\textwidth]{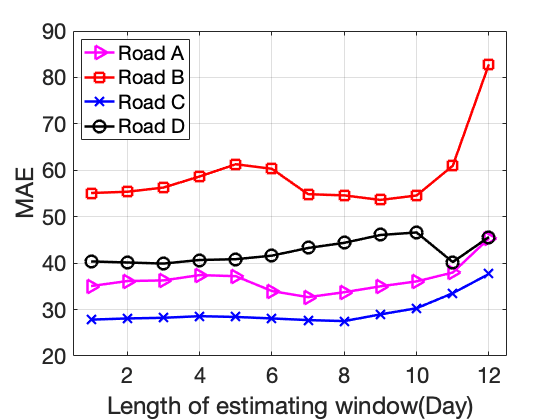}
			\par\vspace{0pt}
		\end{minipage}
	}
    \subfloat [Mean Absolute Percentage Error]
	{
		\label{future_metrics_mape}
		\begin{minipage}[t]{0.3\linewidth}
			\centering
			\includegraphics[width=0.99\textwidth]{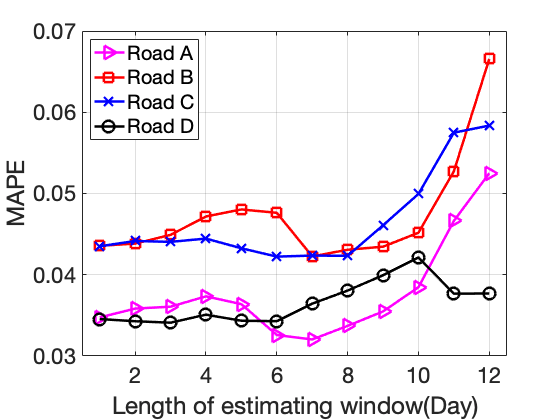}
			\par\vspace{0pt}
		\end{minipage}
	}
	\caption{Prediction accuracy of \BuildSenSys on four roads by varying different lengths of predicting window.}
	\label{future_metrics}
\end{figure*}

We further quantify the performance of \BuildSenSys by comparing it with seven baseline methods by using three evaluation metrics.
The evaluation experiments are conducted on four road segments.
The basic parameter settings of baseline methods are adopted from~\cite{lai2018modeling}.
For \BuildSenSys, we set the length of input time window $L = 24$ and the size of hidden states in both encoder LSTM and decoder LSTM as 128 and 512, respectively.
Table~\ref{baseline_table} presents the performance comparison between~\BuildSenSys and all baseline methods.

Overall, \BuildSenSys achieves the best prediction accuracy with the lowest RMSE at 30.49, lowest MAE at 20.29, and lowest MAPE at 2.05\%, respectively. It outperforms the best prediction of Seq2Seqw/attn by up to
46.2\% in RMSE on road D, 45.6\% in MAE on road C and 65.3\% in MAPE on road D.
In contrast, the LWR model, which also uses building data to predict traffic volume~\cite{zheng2016urban},
shows the worst performance with RMSE at 239.48, MAE at 190.79 and MAPE at 22.13\% on road D.
The key reason here is that the LWR model~\cite{zheng2016urban} is based on the assumption of linear building-traffic correlation.
Meanwhile, the actual building-traffic correlations are proved to be non-linear, time-varying,
and far more complex than the linear model.
For other baseline methods that are based on historical traffic data,
we present the detailed comparison results as follows.

First, ARIMA and VAR methods show the worst prediction performance,
as they are not capable of making accurate predictions for the long term,
especially on the `turning points' (\eg, rush hours).
Interestingly, the Historical Average (HA) method,
as a most naive scheme, performs better than ARIMA and VAR for the following reasons.
Traffic volumes have daily patterns and weekly patterns that are relatively stable in our dataset.
Nevertheless, the HA method can produce unsatisfied prediction results with large errors in holidays,
extreme weather, and social events,
because it can not predict traffic volume that does not follow regular patterns.

Second, the Recurrent Neural Network-based methods,
\ie, LSTM, Seq2Seq, and Seq2Seqw/attn,
show superior performance with MAEs lower than 70 and RMSEs lower than 90.
In specific, the performance of LSTM and Seq2Seq are competitive,
while Seq2Seq outperforms the LSTM by 2.35\% and 2.43\% in MAPE metric on Road D and Road A, respectively.
As for Seq2Seq model based on the encoder-decoder architecture,
it encodes the input traffic volume data into a feature vector,
from which the decoder generates the fixed-length prediction iteratively.
Nevertheless, the performance of encoder-decoder networks will deteriorate rapidly
when the input sequence data becomes longer.
Both \BuildSenSys and Seq2Seqw/attn employ temporal attention mechanisms.
Hence, the decoders of the above models can select the most relevant encoder hidden states adaptively to improve prediction accuracy.
The effectiveness of attention mechanism is validated by the more accurate prediction results by Seq2Seqw/attn and \BuildSenSys.
With 512 hidden states,
both Seq2Seqw/attn and \BuildSenSys greatly improve the prediction accuracy in comparison with models that have 128 hidden states.
More importantly, the prediction accuracy of \BuildSenSys is further enhanced by
the cross-domain attention that learns building-traffic correlations.
In comparison with Seq2Seqw/attn, \BuildSenSys shows 32.3\% and 37.8\% improvements in MAE and RMSE, respectively.
\emph{In summary, compared with RNN-based baseline methods, \BuildSenSys jointly leverages cross-domain attention and temporal attention to adaptively learn cross-domain, time-varying, and non-linear building-traffic correlations.
As a result, \BuildSenSys outperforms all baseline methods with up to 65.3\% accuracy improvement (\eg, 2.2\% MAPE)
in predicting nearby traffic volume.}

\begin{figure*}[t]
	\centering
	\subfloat [Root Mean Squared Error]
	{
		\label{ablation_hiddenstates_rmse}
		\begin{minipage}[t]{0.3\linewidth}
			\centering
			\includegraphics[width=0.99\textwidth]{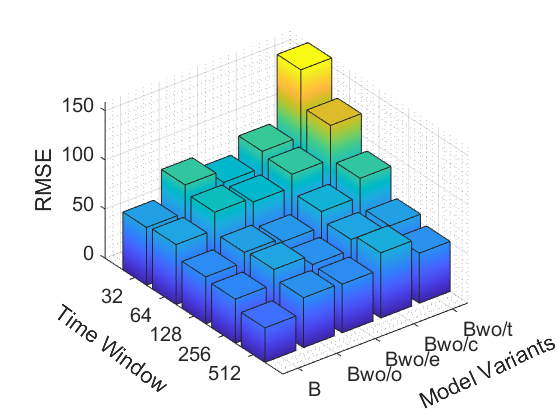}
			\par\vspace{0pt}
		\end{minipage}
	}
	\subfloat [Mean Average Error]
	{
		\label{ablation_hiddenstates_mae}
		\begin{minipage}[t]{0.3\linewidth}
			\centering
			\includegraphics[width=0.99\textwidth]{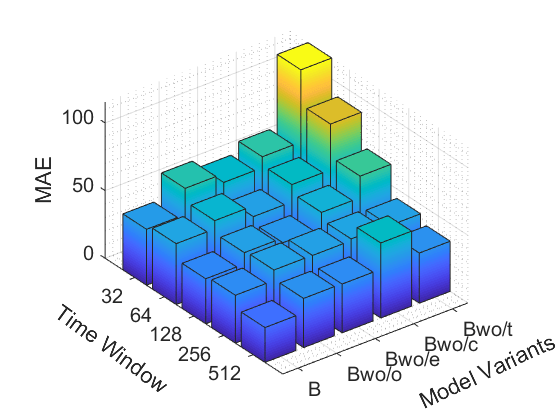}
			\par\vspace{0pt}
		\end{minipage}
	}
    \subfloat [Mean Absolute Percentage Error]
	{
		\label{ablation_hiddenstates_mape}
		\begin{minipage}[t]{0.3\linewidth}
			\centering
			\includegraphics[width=0.99\textwidth]{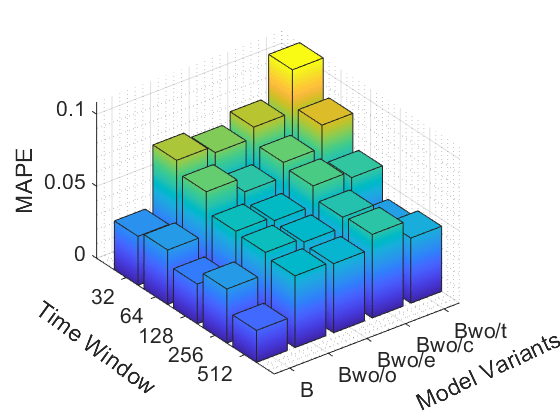}
			\par\vspace{0pt}
		\end{minipage}
	}
	\caption{Performance comparison among different variants of \BuildSenSys with varying size of hidden states.}
	\label{ablation_hiddenstates}
\end{figure*}

\begin{figure*}[t]
	\centering
	\subfloat [Root Mean Squared Error]
	{
		\label{mae}
		\begin{minipage}[t]{0.3\linewidth}
			\centering
			\includegraphics[width=0.99\textwidth]{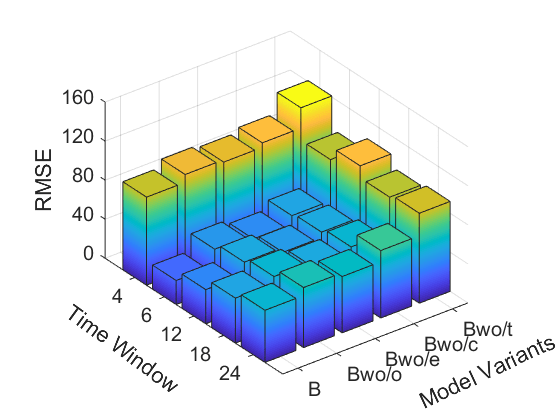}
			\par\vspace{0pt}
		\end{minipage}
	}
	\subfloat [Mean Average Error]
	{
		\label{rmse}
		\begin{minipage}[t]{0.3\linewidth}
			\centering
			\includegraphics[width=0.99\textwidth]{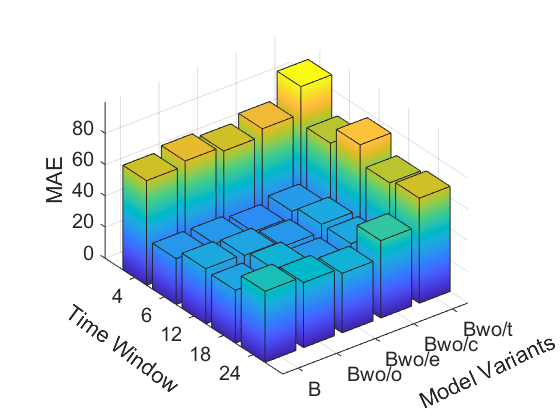}
			\par\vspace{0pt}
		\end{minipage}
	}
    \subfloat [Mean Absolute Percentage Error]
	{
		\label{mape}
		\begin{minipage}[t]{0.3\linewidth}
			\centering
			\includegraphics[width=0.99\textwidth]{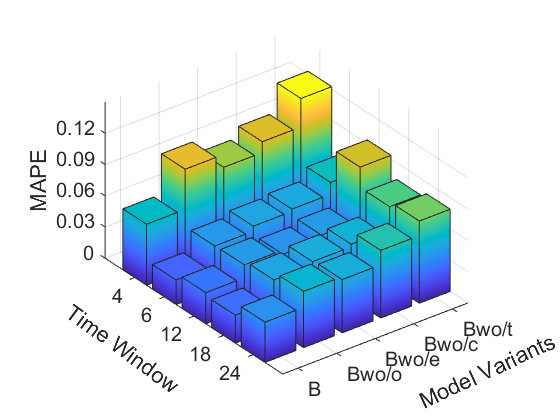}
			\par\vspace{0pt}
		\end{minipage}
	}
	\caption{Performance comparison among different variants of \BuildSenSys with varying length of the time window (in hours).}
	\label{ablation_timewindow}
\end{figure*}

\subsubsection{Evaluation of Parameters}\label{parameter_evaluation}
We further evaluate the impact of parameters in RNN networks on \BuildSenSys's performance,
since \BuildSenSys is based on RNN.
The parameters include the size of the hidden states in the encoder~($h_a$) and decoder~($h_b$),
the lengths of the input time window~($L$) and the prediction window~($\tau$).
Moreover, three RNN-based baseline methods, \ie, LSTM, Seq2Seq, and Seq2Seqw/attn,
are used as the comprised benchmarks.

1) \textbf{Size of hidden states}:
Following~\cite{liang2018geoman}, we set the size of hidden states in both encoder and decoder as the same,
\ie, $h_a=h_b$. We change the value of $h_a$~($h_b$) from 32 to 64, 128, 256, and 512 with a grid search.
As shown in Figs.~\ref{hidden_comparison_rmse}, \ref{hidden_comparison_mae} and \ref{hidden_comparison_mape}, \BuildSenSys outperforms LSTM, Seq2Seq and Seq2Seqw/attn in all settings of hidden states in terms of RMSE, MAE and MAPE.
Moreover, the experimental results show that the prediction accuracy of \BuildSenSys roughly improves with the larger size of hidden states. Overall, \BuildSenSys achieves the best prediction results (\ie, RMSE $\leq 33$, MAE $\leq 25$ and MAPE $\leq 0.22$) when $h_a=h_b=512$.

2) \textbf{Length of input time window}:
The time window $L$ denotes how many hours of historical data are fed into the traffic prediction models (1 hour as the basic unit).
In order to evaluate the impact of $L$, we change its value from 4 to 6, 12, 18, 24 and 48 hours. Thereby, the historical data of the last 4, 6, 12, 18, 24, and 48 hours will be used to predict traffic volume in the next hour.
As illustrated in Figs.~\ref{timewindow_comparison_mae}, \ref{timewindow_comparison_rmse}, and \ref{timewindow_comparison_mape}, the prediction accuracy of all RNN-based methods increases with the greater length of the time window.
Meanwhile, the performance of RNN-based methods also decreases when the time window is too large, \eg,
$L=48$. Accordingly, \BuildSenSys achieves the best performance when $L=6$,
\ie, RMSE$=31.7$, MAE$=24$, and MAPE$=0.019$.

3) \textbf{Length of prediction window:}
We evaluate the prediction accuracy of \BuildSenSys in terms of the length of predicting window~$\tau$.
We change $\tau$ from 1 day to 12 days. 
From Figs.~\ref{future_metrics_rmse}, \ref{future_metrics_mae}, and \ref{future_metrics_mape},
we observe that the RMSE, MAE, and MAPE increase with the greater length of predicting windows,
\ie, the prediction accuracy decreases with the longer predicting windows.
However, except for Road B, the decreasing speed of prediction accuracy is very slow.
Therefore, \BuildSenSys can achieve a stable prediction with high accuracy by reusing building data, \eg,
RMSE and MAE increase from 36 to 52 and 28 to 30, respectively.
Meanwhile, as Road B is the furthest road from the building,
its building-traffic correlation is not as strong as other roads.
Therefore, the prediction errors on Road B increase significantly with the increasing value of $\tau$.

\subsubsection{Evaluation of Cross-domain Learning}
\BuildSenSys incorporates a cross-domain attention mechanism and a temporal attention mechanism to jointly extract spatio-temporal features from building-traffic correlations.
To evaluate the impact of each component on the overall performance,
we conduct an ablation study on \BuildSenSys with its variants as follows.

\begin{itemize}
  \item \textbf{B}, \ie, \BuildSenSys, leverages cross-domain attention and temporal attention to jointly learn building-traffic correlations and temporal correlations with both occupancy data and environmental data.
  \item \textbf{Bwo/o}, a variant of \textbf{B} without \textbf{o}ccupancy component.
  \item \textbf{Bwo/e}, a variant of \textbf{B} without \textbf{e}nvironmental component.
  \item \textbf{Bwo/c}, a variant of \textbf{B} without \textbf{c}ross-domain attention, which only employs temporal attention to learn temporal dependencies of building sensing data and traffic volume data.
  \item \textbf{Bwo/t}, a variant of \textbf{B} without \textbf{t}emporal attention, which only employs cross-domain attention to learn cross-domain building-traffic correlations.
\end{itemize}

First, we evaluate the performance of different components of \BuildSenSys by changing the size of hidden states.
As shown in Fig.~\ref{ablation_hiddenstates}, \BuildSenSys demonstrates the best performance,
while Bwo/t shows the worst performance as it lacks temporal correlations for traffic prediction.
Meanwhile, the prediction accuracy of both Bwo/o and Bwo/e is better than Bwo/c.
It indicates that cross-domain learning with building data can effectively improve overall performance. Moreover, as illustrated in Figs.~\ref{ablation_hiddenstates_rmse}, \ref{ablation_hiddenstates_mae}. and \ref{ablation_hiddenstates_mape},
the prediction accuracy of Bwo/e is higher than that of Bwo/o. This result shows that the building occupancy data has more contribution to cross-domain traffic prediction than building environmental data.

Second, we evaluate the performance of different components of \BuildSenSys by changing the length of the input time window.
As shown in Fig.~\ref{ablation_timewindow},
the performance of Bwo/t deteriorates significantly as it has no temporal attention to capture the temporal correlations in building data and traffic data.
As a result, the prediction accuracy of Bwo/t is the worst,
\eg, the RMSE over 115, MAE over 80, and MAPE over 10\%, respectively.
Meanwhile, \BuildSenSys shows the best prediction accuracy
when the input time window is 6.
However, when the length of the input time window increases to 12, 18, and 24,
the prediction accuracy of all variants decreases continuously.
This can be explained as the longer inputs could pose greater difficulty in capturing building-traffic correlations and temporal correlations.

To sum up, the above results indicate that each component in cross-domain learning has its contribution to enhancing the prediction accuracy of \BuildSenSys. Specifically, the occupancy component and the environmental component are complementary to each other in reusing building data to predict nearby traffic volume. Moreover,
it is highly significant to jointly leverage cross-domain attention and temporal attention in cross-domain learning of building data.
Each attention mechanism has its unique contribution to the improvement of prediction accuracy, \eg, up to 45.5\% improvement by temporal attention and 30.9\% improvement by cross-domain attention.

\subsubsection{Evaluation of Attention Weight}\label{attention_analysis}
\begin{figure}[t]
\centering
\subfloat [Cross-domain Attention Weight]
	{
		\label{cross-domain_attn}
		\begin{minipage}[t]{0.48\linewidth}
			\centering
			\includegraphics[width=0.99\textwidth]{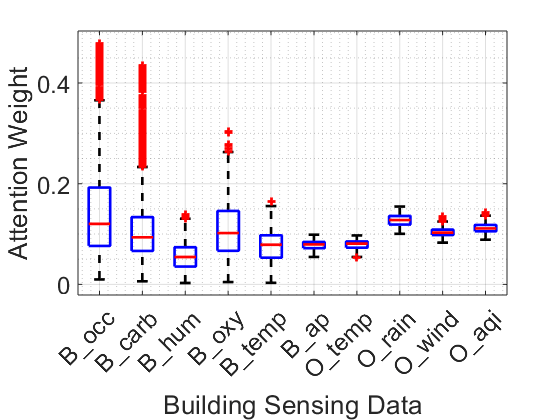}
			\par\vspace{0pt}
		\end{minipage}
	}
\subfloat [Temporal Attention Weight]
	{
		\label{temporal_attn} 
		\begin{minipage}[t]{0.48\linewidth}
			\centering
			\includegraphics[width=0.99\textwidth]{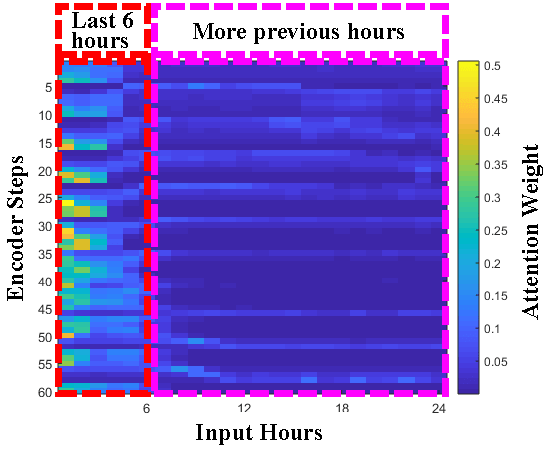}
			\par\vspace{0pt}
		\end{minipage}
	}
	\caption{Visualization of Cross-domain Attention and Temporal Attention.}
	\label{attn_visual} 
\end{figure}

\begin{figure*}[t]
\centering
\subfloat [Distinct average daily traffic.]
	{
		\label{daily_average}
		\begin{minipage}[t]{0.33\linewidth}
			\centering
			\includegraphics[width=0.99\textwidth]{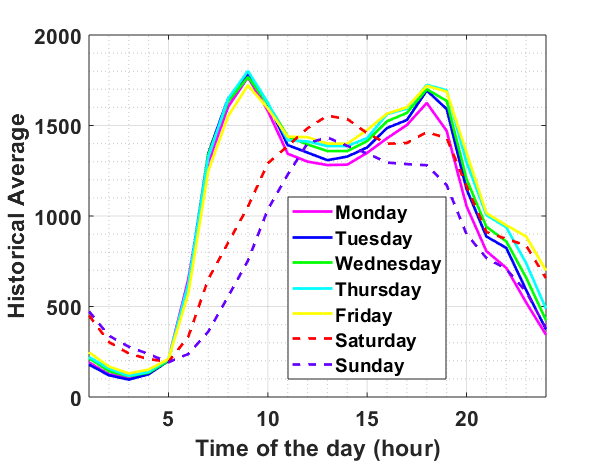}
			\par\vspace{0pt}
		\end{minipage}
	}
\subfloat [Relative prediction error by different models.]
	{
		\label{two_models} 
		\begin{minipage}[t]{0.33\linewidth}
			\centering
			\includegraphics[width=0.99\textwidth]{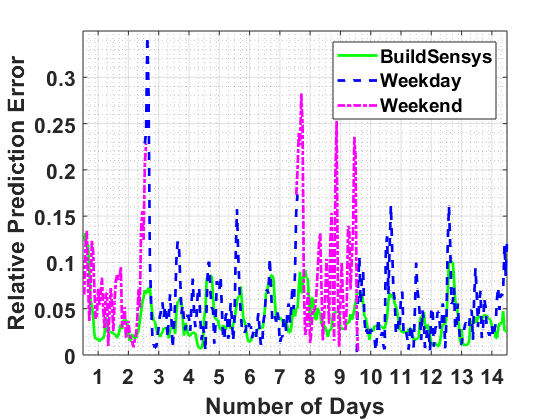}
			\par\vspace{0pt}
		\end{minipage}
	}
\subfloat [Comparison with state-of-the-art models.]
	{
		\label{stateoftheart} 
		\begin{minipage}[t]{0.33\linewidth}
			\centering
			\includegraphics[width=0.99\textwidth]{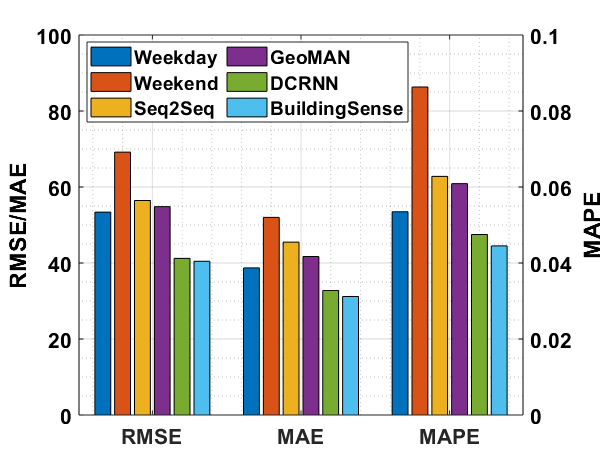}
			\par\vspace{0pt}
		\end{minipage}
	}
	\caption{Performance comparison with the Weekday model and Weekend model.}
	\label{new_comparison} 
\end{figure*}

1) {\bf{Cross-domain attention}}: We analyze the attention weights of the proposed traffic prediction model in two folds, \ie, cross-domain attention weight and temporal attention weight. First, as the correlations between traffic volume data and different building sensing data are non-linear, \BuildSenSys employs a cross-domain attention mechanism to achieve the non-linear mapping from building sensing data to traffic data in predicting traffic volume.
To demonstrate the different attention weights from different types of building sensing data,
we conduct extensive experiments to analyze the attention weight of each type of building sensing data. As shown in Fig.~\ref{cross-domain_attn}, the 10 groups of attention weights correspond to 10 types of sensing data, including building occupancy (B\_occ), ${\rm{C}}{{\rm{O}}_2}$ concentration (B\_carb), building humidity (B\_hum), ${{\rm{O}}_2}$ concentration (B\_oxy),
building temperature (B\_temp), building air pollution (B\_ap), as well as outdoor temperature (O\_temp), rainfall (O\_rain) and air quality index (O\_aqi).
The box plot in Fig.~\ref{cross-domain_attn} compares the distributions of attention weights
by each type of building sensing data.


As demonstrated in Fig.~\ref{cross-domain_attn}, each type of building sensing data has its own contribution to traffic prediction while building occupancy data has the highest attention weights across all predicting steps. Specifically, the whiskers (\ie, lines extending above and below each box) of B\_occ have the most extensive range (from 0.01 to 0.36).
B\_occ covers the furthest adjacent values of attention weight, showing the significant importance of building occupancy data in cross-domain traffic prediction.
Moreover, the B\_occ and B\_carb have the largest number of outliers with the highest values (\ie, up to 0.48 and 0.39, respectively). It indicates that they have stronger correlations with traffic data and higher attention weights in traffic prediction. Finally, the median values (\ie, the central marks) of attention weights boxes are roughly close, illustrating that all types of building data make non-negligible contributions in traffic predictions.

In summary, the above experimental results indicate that,
each building data has its own contribution to enhancing the prediction accuracy of \BuildSenSys.
Meanwhile, the building occupancy data and the environmental data are complementary to each other in reusing building data for traffic sensing and prediction.

2) {\bf{Temporal attention}}: We further investigate how the temporal attention mechanism captures the correlations between predicted traffic values and historical traffic data.
More specifically, we conduct the exploring experiment to visualize how \BuildSenSys assigns different attention weights to the historical traffic data in the time window `L'.
As explained in the experimental setups,
the input time window `L' represents the length of input data measured in hours,
and we set `L' as 24 (hours) in the following experiment.

As a result, the heat map in Fig.~\ref{temporal_attn} visualize the temporal attention weights over 24 hours and 60 predicting steps.
We divide the input time window into two categories, \ie, the latest 6 hours and more previous (last 6-24) hours. The above two categories have great disparities in assigned attention weights.
For the input data of the latest 6 hours, \BuildSenSys assigns more attention weight to them,
as each hour's data has up to 0.5 attention weight.
In contrast, most input data in the last 6-24 hours have particularly small values of attention weights (under 0.05).
The above result shows that the temporal attention mechanism in \BuildSenSys would focus more on the latest 6 hours of the input data.
Meanwhile, as all of the temporal attention weights sum to 1,
with the longer input time windows (\eg, 12, 18, 24 and 48 in our ablation studies),
the latest 6 hours will share less attention weight.
In such cases, \BuildSenSys will not be able to specifically focus on the most recent data,
but to process both the short-term and long-term temporal correlations for traffic volume prediction.
Consequently, the prediction accuracy of \BuildSenSys will be influenced,
as demonstrated and verified by the evaluation results in previous experimental studies.

\subsubsection{Extensive evaluations of comparison}\label{revision_experiments}
\textbf{Baseline methods with the weekday's data and the weekend's data:}
To further explore the impact of data patterns on the traffic prediction results, we conduct an extensive experiment by applying two different models to predict traffic volumes on weekdays and weekends, respectively.
First, we visualize a distinct daily average of traffic data on a target road in Fig.\ref{daily_average},
where different colors plot the average values of one-year traffic data on each day of the week.
We can observe that the traffic volumes on weekdays and weekends follow two different patterns, respectively.
For weekdays, the peak hours range from 8 a.m. to 10 a.m. and 4 p.m. to 7 p.m.
Meanwhile, for weekends, there is a longer range for peak hours, i.e., 12 a.m. to 6 p.m.
We aim to study whether using two different models of \BuildSenSys would impact the results of overall results of traffic volume prediction.

The basic settings for this experiment are the same as experiments in 7. The only difference is that we divide one-year traffic data into two groups of data (weekdays and weekends), and correspondingly we train two models (weekday model and weekend model) of BuildSenSys for each group. At the same time, we train a BuildSenSys model based on the whole dataset, and the total epochs for all models are set as 2,500. The prediction results of one-model prediction and two-model prediction are shown in Fig.~\ref{two_models} and Fig.~\ref{stateoftheart}. Weekday represents a special model of \BuildSenSys that is trained only with weekday data to predict the data on weekdays. Similarly, Weekend stands for another special model of \BuildSenSys that is trained only with weekend data predicts the data on weekends.
As illustrated in Fig.~\ref{two_models} and Fig.~\ref{stateoftheart},
we have compared the prediction accuracy of the Weekday model and the Weekend model, Seq2Seq model, and \BuildSenSys model. The results show that in comparison with the Seq2Seq model, both the Weekday model and the Weekend model can achieve some improvements in prediction accuracy. However, the combined prediction results are still not satisfactory with even lower accuracy than the Weekday model.
Meanwhile, by using a temporal attention mechanism, the proposed \BuildSenSys can dynamically learn the impact of historical data on the predicting target, thereby it outperforms both the Weekday model and the Weekend model. The experimental results also validate that, breaking the continuity of the dataset in the temporal dimension will undermine the performance of temporal attention mechanisms.

\textbf{Comparison with state-of-the-art methods based on different data sources:}
To address how different data measurements would impact the traffic prediction results, we further employ three baseline methods for comparison. Among these baseline methods, the Sequence to Sequence model (Seq2Seq)~\cite{sutskever2014sequence} uses traffic volume data from traffic sensing system on one road for prediction;
the Multi-level Attention Networks for Geospatial Sensors (GeoMAN)~\cite{liang2018geoman} exploits the data of one road along with the weather data; and Diffusion Convolutional Recurrent Neural Network (DCRNN)~\cite{li2018diffusion} leverages the sensing data from multiple roads with the road graph. In this work, BuildSenSys leverages building sensing data to predict the traffic volume on nearby roads.
The experimental results in Fig.~\ref{stateoftheart} show that BuildSenSys achieves the best performance, even compared with DCRNN,
which uses traffic data from 10 road segments to predict traffic volume on the target road.
The GeoMAN outperforms Seq2Seq,
as it considers both spatial and temporal attention with weather data and historical road data for traffic prediction.
Meanwhile, DCRNN further improves the prediction accuracy by using both diffusion convolution and sequence to sequence learning framework on traffic volume data of multiple roads with sensor topology information.
The above results also assess the practicability and usefulness of cross-domain traffic prediction by re-using building sensing data.

\section{discussion}\label{discussion}

In this work, we explore the possibility of reusing building sensing data to predict the traffic volume of nearby road segments.
Although we have made some advances toward this new direction,
there are still several issues that need to be further investigated as follows.

\textbf{Applicable conditions of \BuildSenSys}:
It is of great importance to discuss: 1) which kind of buildings are suitable for accurate traffic sensing and prediction, and 2) what kind of real-world applications could benefit from cross-domain traffic prediction by reusing building sensing data.
Based on our initial studies, we roughly summarize three critical factors of a building for cross-domain traffic prediction, including location, capacity and building types.
First, human movement patterns (including traffic) among urban sites are affected by buildings,
which can `temporarily hold' human mobility~\cite{zheng2018buildings}.
Therefore, buildings located close to road networks are preferable for traffic sensing,
as they would have more chances to affect human mobility on the road ~\cite{zheng2018buildings}.
Second, the capacity of a building, \ie, the volume of occupants it can hold,
is essential for accurate cross-domain traffic sensing.
For instance, Zheng~\etal~\cite{zheng2016urban} showed a commercial building with a capacity of 10000 occupants can effectually impact the traffic on nearby roads with substantial evidence.
Third, the types of buildings can affect the applicable conditions of \BuildSenSys.
Different types of buildings have different patterns of occupancy dynamics,
which will directly affect building-traffic correlations.
To this end, we recommend commercial building (\eg, office buildings and retail buildings) are the first choice when performing cross-domain traffic sensing with \BuildSenSys.
In practice, \BuildSenSys can be used in a variety of real-world applications. For example, it can be exploited to provide real-time traffic volume data to support controls of intelligent traffic light~\cite{zhang2019cityflow, wei2018intellilight}. Besides, \BuildSenSys can also be applied in traffic management systems, which requires accurate traffic predictions,
especially in rush hours~\cite{zhu2018big}.

\begin{figure}[t]
	\centering
	\includegraphics[width=0.40\textwidth]{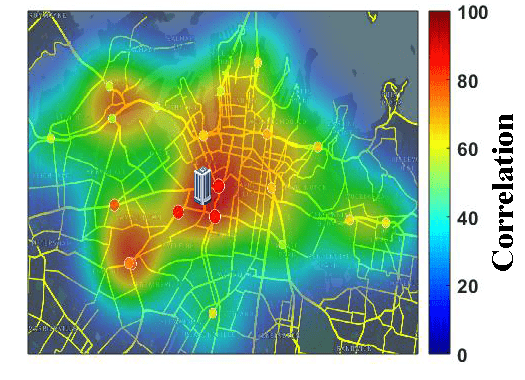}
	\caption{Illustrations of prediction coverage by \BuildSenSys.}
	\label{sensing_coverage}
\end{figure}

\textbf{Sensing coverage of \BuildSenSys:}
In this paper, \BuildSenSys is a proof-of-concept in reusing building data for traffic sensing. As shown in the experimental evaluations of Section 6, the prediction accuracy  of \BuildSenSys is in accordance with the building-traffic correlation of a road.
Thus, we further conduct experiments to investigate the coverage of cross-domain traffic prediction by reusing building sensing data. In specific, as illustrated in Fig.~\ref{sensing_coverage}, we calculate the correlations between building occupancy and traffic data on nearby roads using the Pearson Correlation Coefficient, and further use the building-traffic correlations as the index of sensing coverage of \BuildSenSys. Fig.~\ref{sensing_coverage} demonstrates that BuildSenSys generally has a prediction coverage within 5 km to the building.

\textbf{Extension to large-scale scenarios:}
According to the investigation of sensing coverage,
the \BuildSenSys system can predict traffic volume for road segments within 5 km of the building.
To extend this coverage of traffic predictions to a larger scale,
it requires to incorporate more sensing data that are correlated with traffic volume.
Intuitively, traffic data of road segments in the same district would have spatial and temporal correlations~\cite{zhu2009seer,liu2016mining}.
For example,
Liu~\etal~\cite{liu2017participatory,liu2018think} probed traffic conditions on different road segments with GPS data from bus riders.
In our future work, we will further exploit the above multi-road and multi-region correlations to extend the coverage of \BuildSenSys in traffic sensing and prediction.
Besides, the pervasive street cameras are the cost-efficient data source for aggregating urban crowd flow, which may have direct and indirect correlations with traffic volume on nearby roads.
Existing works in computer vision have proposed highly effective approaches to generate density maps~\cite{xu2019learn} and achieve accurate crowd counting~\cite{cao2018scale},
which have the potential to contribute to cross-domain traffic sensing and prediction.

\textbf{Privacy and security of building data:}
With the massive amount of IoT data generated by smart buildings,
it is essential to reuse building data in a privacy-preserving manner.
In our work, the building datasets, including occupancy data and environmental data,
are anonymized by the data provider.
As a consequence, they cannot be used to trace back to any personal information.
Notably, the people counters (cameras) do not have facial recognition capability and
there would be no personal privacy leakage from the building occupancy data.
Moreover, if needed, we can adapt existing privacy protection methods~\cite{jia2017privacy,zhang2017security} to prevent privacy leakage when utilizing building sensing data.

\section{Conclusion}\label{conclusion}
In this paper, we have proposed \BuildSenSys,
a first-of-its-kind building sensing data-based traffic volume prediction system with cross-domain learning.
Firstly, we have conducted extensive experimental analysis on building-traffic correlations based on multi-source real-world datasets.
It discloses that building data has strong correlations with traffic data.
Then, we have proposed a cross-domain learning-based RNN with cross-domain and temporal attention mechanisms to jointly extract building-traffic correlations for accurate traffic prediction.
Moreover, we have implemented a prototype system of \BuildSenSys and conducted extensive experiments.
The experimental results demonstrated that \BuildSenSys outperforms seven baseline methods with
up to 65.3\% accuracy improvement in predicting nearby traffic volume.
We believe this work can open a new gate to reuse building sensing data for traffic sensing and prediction,
hence indicating an interconnection between smart buildings and intelligent transportation.

\section*{Acknowledgments}
The authors would like to thank Zhao Zhang, Xiang Huang, Qingqing Wang, Xiaopu Zhang and Xianghao Chu for their support and assistance with this work. The authors are grateful to the anonymous reviewers for their valuable comments. This research is supported by the NSF of China Projects: Grants No.61632010, 61872447, 61772546, 61902211, the Natural Science Foundation of Chongqing: Grant No.CSTC2018JCYJA1879, and China Scholarship Council: Grant No.201603170125.

\bibliographystyle{IEEEtran}
\bibliography{main}

\end{document}